\documentclass[aps,prd,preprint,onecolumn,groupedaddress,nofootinbib]{revtex4}

\usepackage{latexsym}
\usepackage{amsthm}
\usepackage{amsmath}
\usepackage{graphicx}
\usepackage{subcaption}
\usepackage{mathtools}
\usepackage{slashed}
\usepackage{amssymb}
\usepackage{indentfirst}
\usepackage{color}
\usepackage{epsf}
\definecolor{mycolor}{RGB}{13,5,255}
\usepackage[colorlinks=True, citecolor=blue, urlcolor=blue, linkcolor=blue]{hyperref}

\newcommand*\diff{\mathop{}\!\mathrm{d}}
\newcommand*\Diff[1]{\mathop{}\!\mathrm{d^#1}}
\newcommand{\nn}{\nonumber}

\newcommand{\be}{\begin{eqnarray}}
\newcommand{\ee}{\end{eqnarray}}
\newcommand{\ma}{\mathrm}
\newcommand{\ml}{\mathcal}
\newcommand{\bs}{\boldsymbol}
\newcommand{\Tr}{\mathrm{Tr}}

\begin{document}

\title{Quarkonium inside Quark-Gluon Plasma: Diffusion, Dissociation, Recombination and Energy Loss}

\author{Xiaojun Yao}
\email{xiaojun.yao@duke.edu}
\author{Berndt M\"uller}
\email{mueller@phy.duke.edu}
\affiliation{Department of Physics, Duke University, Durham, NC 27708, USA}

\date{\today}

\begin{abstract}
We consider the quarkonium diffusion, dissociation and recombination inside quark-gluon plasma. We compute scattering amplitudes in potential nonrelativistic QCD for relevant processes. These processes include the gluon absorption/emission at the order $gr$, inelastic scattering at the order $g^2r$ and elastic scattering with medium constituents at the order $g^2r^2$. We show these amplitudes satisfy the Ward identity. We also consider one-loop corrections. The dipole interaction between the color singlet and octet is not running at the one-loop level. Interference between the tree-level gluon absorption/emission and its thermal loop corrections cancels the collinear divergence in the $t$-channel inelastic scattering. The inelastic scattering has no soft divergence because of the finite binding energy of quarkonium. We write out the diffusion, dissociation and recombination terms explicitly for a Boltzmann transport equation and define the dissociation and recombination rates. Furthermore, we calculate the diffusion coefficient of quarkonium. We find our result of diffusion coefficient differs from a previous calculation by two to three orders of magnitude. We explain this and can reproduce the previous result in a certain limit. Finally we discuss two mechanisms of quarkonium energy loss inside quark-gluon plasma. 
\end{abstract}

\maketitle

\section{Introduction}
\label{sect:intro}
Since the early study of static screening on quarkonium \cite{Matsui:1986dk}, the bound state of a heavy quark antiquark pair, quarkonium has been used as a probe of quark-gluon plasma (QGP) in heavy ion collisions. In a high temperature QGP, the attractive potential between the heavy quark antiquark  pair $Q\bar{Q}$ is significantly suppressed and thus the bound state cannot exist. The screening effects have been widely investigated by computing the free energy of $Q\bar{Q}$ \cite{McLerran:1981pb,Kaczmarek:2002mc} or spectral functions \cite{Asakawa:2003re,Datta:2003ww} on a lattice at finite temperature. 

In addition to the static screening effect, the dynamical screening effect also exists inside QGP. It is the dissociation of quarkonium caused by collisions with medium constituents. The dissociation process at leading order (LO) in the coupling constant is the gluon absorption process $g+H\rightarrow Q+\bar{Q}$ where $H$ indicates a quarkonium state. It was first investigated by using large-$N_c$ expansions \cite{Peskin:1979va,Bhanot:1979vb}. At next-leading order (NLO), inelastic scattering between quarkonium and medium constituents contributes to the dissociation $l(\bar{l},g)+H\rightarrow l(\bar{l},g)+Q+\bar{Q}$ where $l$ denotes a light quark. The inelastic scattering was first studied in the quasi-free limit where the $Q$ and $\bar{Q}$ are treated as free  particles and each of them scatters independently with medium constituents \cite{Grandchamp:2001pf}. Later, the interference effect was taken into account. This leads to a dependence of the dissociation rate on the relative position of the heavy quark antiquark pair \cite{Laine:2006ns,Beraudo:2007ky}. This maps into a dependence of the inelastic scattering on the bound state wave function \cite{Park:2007zza}, as in the case of gluon absorption. The dissociation rate can be interpreted as an imaginary part of the $Q\bar{Q}$ potential. One can also interpret the dissociation as a decoherence of the quarkonium wave function in the language of open quantum systems \cite{Akamatsu:2011se}. More recently, these dissociation rates were studied from the thermal loop corrections of the singlet propagator in potential nonrelativistic QCD (pNRQCD) \cite{Brambilla:2010vq,Brambilla:2011sg,Brambilla:2013dpa} by systematic weak coupling and nonrelativistic expansions. 
Anisotropic corrections to dissociation rates have also been considered \cite{Dumitru:2007hy, Dumitru:2009fy, Du:2016wdx}.

To describe the transport of quarkonium inside QGP, one also needs to consider the in-medium recombination from unbound $Q\bar{Q}$ pairs \cite{Thews:2000rj}. This can be modeled by detailed balance and a phenomenological factor controlling how much open heavy quarks are thermalized \cite{Du:2017qkv}. Recombination from parametrized nonequilibrium heavy quark distributions has also been investigated \cite{Song:2012at}. For practice, one needs heavy quark distributions from real in-medium dynamics. To this end, one can couple the transport equations of quarkonium with those of heavy quarks \cite{Yao:2017fuc}. The inverse of the gluon absorption, the gluon emission, $Q+\bar{Q} \rightarrow g+H $ has been simulated in Ref.~\cite{Yao:2017fuc} with dynamically evolving heavy quark distributions and the approach to detailed balance and equilibrium was demonstrated.

Besides dissociation and recombination, quarkonium can also diffuse inside QGP because it is approximately a color dipole, and thus not exactly color neutral. It may elastically scatter with medium constituents. The diffusion coefficient of quarkonium has been estimated in both weak and strong coupling limits \cite{Dusling:2008tg}. However, the diffusion has not been included in phenomenological studies using transport equations.

In this paper, we consider the dissociation, recombination and diffusion of quarkonium in the same theoretical framework. We apply pNRQCD to calculate the relevant terms in a Boltzmann transport equation. The transport equation can be derived from first principles by using the open quantum system formalism and effective field theory pNRQCD \cite{Yao:2018nmy}.
The use of open quantum system formalism to study quarkonium in-medium dynamics has been widely investigated recently \cite{Akamatsu:2014qsa,Blaizot:2017ypk,Blaizot:2018oev,Brambilla:2016wgg,Brambilla:2017zei}.
We compute directly the scattering amplitudes of gluon absorption/emission, inelastic scattering and elastic scattering in pNRQCD, in contract to previous studies of dissociation rates that use the optical theorem and loop corrections to the forward amplitudes \cite{Brambilla:2010vq,Brambilla:2011sg,Brambilla:2013dpa}. By writing out the amplitudes explicitly, we can show these amplitudes satisfy the Ward identities, which has not been explicitly shown before. We also consider the loop correction to the gluon absorption/emission. We demonstrate that the $t$-channel inelastic scattering is infrared safe. Furthermore, we compute the diffusion coefficient of quarkonium and reproduce a previous result in a certain limit. Finally, we discuss two mechanisms of quarkonium energy loss inside QGP: one through diffusion and the other via dissociation first and then recombination later.

The paper is organized as follows. In Section~\ref{sect:boltzmann}, we briefly introduce the Boltzmann transport equation of quarkonium. Then in Section~\ref{sect:pnrqcd}, we explain the effective field theory pNRQCD used in the calculations. The calculation results of dissociation, recombination and diffusion are shown in the following Sections~\ref{sect:disso_reco} and \ref{sect:diffuse}. In Section~\ref{sect:diffuse}, two mechanisms of quarkonium energy loss are also discussed. Finally the conclusions are drawn in Section~\ref{sect:conclusion}.

\section{Boltzmann transport equations}
\label{sect:boltzmann}
The dynamical evolution of quarkonium inside QGP can be described by Boltzmann transport equations for the distribution function of each quarkonium state with the quantum number $nl$ and spin $s$,
\be
\label{eqn:boltzmann}
\frac{\partial}{\partial t} f_{nls}({\bs x}, {\bs p}, t) + {\bs v}\cdot \nabla_{\bs x}f_{nls}({\bs x}, {\bs p}, t) = \ml{C}_{nls}^{+}({\bs x}, {\bs p}, t) - \ml{C}_{nls}^{-}({\bs x}, {\bs p}, t) + \ml{C}_{nls}({\bs x}, {\bs p}, t)\,.
\ee
The three collision terms $\ml{C}_{nls}^{+}$, $\ml{C}_{nls}^{-}$ and $\ml{C}_{nls}$ represent the recombination, dissociation and diffusion of quarkonium in the medium respectively. In the following when we compute the square of the scattering amplitudes, we will average over the polarizations of non-S wave quarkonium states. So we omit the quantum number $m$ throughout the paper. 
The Boltzmann transport equation can be derived from QCD by using the open quantum system formalism and effective field theory under the assumption of Markovian process and weak coupling between the quarkonium and the medium \cite{Yao:2018nmy}. When the quarkonium size is smaller than the screening length of the medium, the weak coupling assumption is valid. Both the nonrelativisitic and weak coupling expansion parameters are smaller for bottomonium than charmonium. This implies that the leading order results are more reliable for bottomonium. 

In the following sections, we will compute these collision terms in pNRQCD.

\section{Potential NRQCD}
\label{sect:pnrqcd}
The effective field theory pNRQCD can be systematically derived from QCD under the separation of scales: $M\gg Mv \gg Mv^2$ \cite{Brambilla:1999xf}. Here $M$ denotes the mass of heavy quark, assumed to be large and $v$ is the typical relative velocity between the heavy quark antiquark pair inside a quarkonium. For charmonium, $v^2\sim0.3$ while for bottomonium, $v^2\sim0.1$ \cite{Bodwin:1994jh}. To derive pNRQCD, one integrates out the scales $M$ and $Mv$ in sequence, and applies a double expansion in the heavy quark velocity $v$ (nonrelativistic expansion) and the inter-quark distance $r\sim\frac{1}{Mv}$ (multipole expansion). For quarkonium inside QGP, two additional scales appear: the plasma temperature $T$ and Debye mass $m_D$. Depending on where $T$ and $m_D$ fit into the vacuum separation of scales $M\gg Mv \gg Mv^2$, one can have different versions of the theory. Here we focus on the following hierarchy of scales: $M\gg Mv \gg Mv^2\gtrsim T\gtrsim m_D$ where $T$ and $m_D$ are on the same order as the typical binding energy $Mv^2$. In this hierarchy, a bound quarkonium state can be well-defined. If $T$ and $m_D$ are on the order of $Mv$, the Debye screening will be so strong that the potential cannot support any bound state. In cases with $rT\sim rm_D\sim1$, the multipole expansion also breaks down.

The Lagrangian density of pNRQCD is given by
\be
\label{eq:lagr}
\ml{L}_\ma{pNRQCD} &=& \int \diff^3r \Tr\Big(  \ma{S}^{\dagger}(i\partial_0-H_s)\ma{S} +\ma{O}^{\dagger}( iD_0-H_o )\ma{O} + V_A( \ma{O}^{\dagger}\bs r \cdot g{\bs E} \ma{S} + \ma{h.c.})  \\\nn
&&+ \frac{V_B}{2}\ma{O}^{\dagger}\{ \bs r\cdot g\bs E, \ma{O}  \} +\cdots \Big) + \ml{L}_\ma{light\ quark} + \ml{L}_\ma{gluon} \, ,
\ee
where ${\bs E}$ represents the chromoelectric field and $D_0\ma{O} = \partial_0\ma{O} -ig [A_0, \ma{O}]$. The gluon and light quark parts are just QCD with momenta $\lesssim Mv$. The degrees of freedom are the color singlet $\ma{S}(\bs R, \bs r, t)$ and color octet $\ma{O}(\bs R, \bs r, t)$ where $\bs R$ denotes the center-of-mass (c.m.) position and $\bs r$ the relative coordinate. We will assume the medium is translationally invariant so the existence of the medium does not break the separation into the c.m.~and relative motions. The color singlet and octet Hamiltonians are expanded in powers of $1/M$:
\be
H_{s} &=& \frac{(i\bs \nabla_\ma{cm})^2}{4M} + \frac{(i\bs \nabla_\ma{rel})^2}{M} + V_{s}^{(0)} + \frac{V_{s}^{(1)}}{M} + \frac{V_{s}^{(2)}}{M^2} + \cdots\\
H_{o} &=& \frac{(i\bs D_\ma{cm})^2}{4M} + \frac{(i\bs \nabla_\ma{rel})^2}{M} + V_{o}^{(0)} + \frac{V_{o}^{(1)}}{M} + \frac{V_{o}^{(2)}}{M^2} + \cdots\,.
\ee
We will work to the lowest order in the expansion of $v$. By the virial theorem, $\bs p_\ma{rel}^2/M \sim V_{s,o}^{(0)}\sim Mv^2$. Higher-order terms of the potentials including the relativistic corrections and spin-orbital and spin-spin interactions are suppressed by extra powers of $v$. In the $Q\bar{Q}$ pair (bound or unbound) rest frame, the initial c.m.~momentum is zero. If the medium is static with respect to the $Q\bar{Q}$ pair, the final c.m.~momentum after a scattering is of order $\sim T$. Since in our power counting, $T\lesssim Mv^2$, the c.m.~kinetic energy is of order $\lesssim Mv^4$ and thus suppressed by $v^2$.\footnote{If the medium is moving with respect to the $Q\bar{Q}$ pair at a velocity $v_\ma{med}$, the c.m.~kinetic energy is still suppressed at least by one power of $v$ if $v_\ma{med}\lesssim\sqrt{1-v}$. We assume the medium is static with respect to the $Q\bar{Q}$ pair in this paper. Generalization to the case of moving medium with $v_\ma{med}\lesssim\sqrt{1-v}$ can be easily worked out.} Therefore,
\be
\label{eqn:hamiltonian}
H_{s,o}  =  \frac{(i\bs \nabla_\ma{rel})^2}{M} + V_{s,o}^{(0)}\,.
\ee
The potentials and Wilson coefficients $V_{A,B}$ in the chromoelectric dipole vertices can be obtained by matching pNRQCD with NRQCD \cite{Brambilla:1999xf,Brambilla:2004jw,Fleming:2005pd}. Up to order $g^2r$, 
\be
\label{eqn:match}
V_{s}^{(0)} = -C_F\frac{\alpha_s}{r}\,,\ \ \ \ \ \ \ V_{o}^{(0)} = \frac{1}{2N_c}\frac{\alpha_s}{r}\,,\ \ \ \ \ \ \ V_A=V_B=1\,.
\ee
The chromomagnetic vertices are suppressed by powers of $v$. The potential is Coulomb, which is approximately valid inside QGP since the confining part is flattened. One can improve the potentials by doing nonperturbative matching calculations at finite temperature.

Under a gauge transformation $U(\bs R, t)$,
\be
\ma{S}(\bs R, \bs r, t) &\rightarrow& \ma{S}(\bs R, \bs r, t)\\
\ma{O}(\bs R, \bs r, t) &\rightarrow& U(\bs R, t)\ma{O}(\bs R, \bs r, t)U^\dagger(\bs R, t)\\
D_\ma{cm} ^\mu &\rightarrow&    U(\bs R, t)D_\ma{cm} ^\mu U^\dagger(\bs R, t)\,,
\ee
therefore the Lagrangian density is invariant under a gauge transformation associated with the c.m.~motion. It is worth noting that the relative motion is not gauged due to the multipole expansion. The potentials may be gauge-dependent at the matching calculation. 

\begin{figure}
	\centering
	\begin{tabular}{  c  c  l  }
	\hspace*{-0.26in}{ \raisebox{-0.26in}{\includegraphics[height=0.35in]{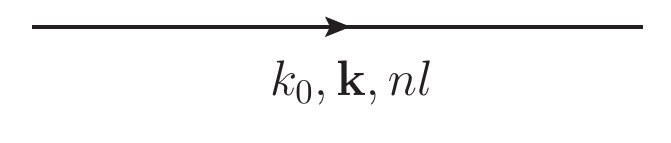}} } &=\ \ \ \ & $\frac{i| \psi_{nl} \rangle \langle \psi_{nl} | }{k_0-\frac{{\bs k}^2}{4M} + |E_{nl}| +i\epsilon } $\\
	\raisebox{-0.26in}{\includegraphics[height=0.5in]{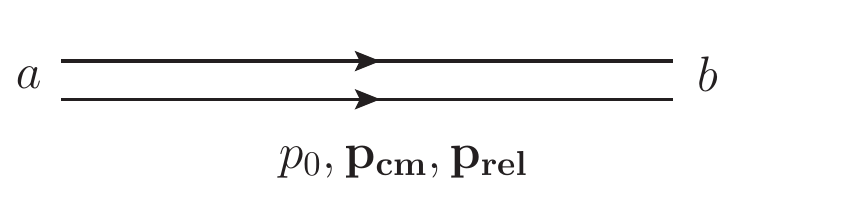}}   &=\ \ \ \ & $\frac{i| \Psi_{{\bs p}_{\ma{rel}}} \rangle \langle \Psi_{{\bs p}_{\ma{rel}}} |}{p_0-\frac{{\bs p^2_{\ma{cm}}}}{4M} - \frac{{\bs p^2_{\ma{rel}}}}{M} +i\epsilon } \delta^{ab}$ \\
	\raisebox{-0.26in}{ \includegraphics[height=1.3in]{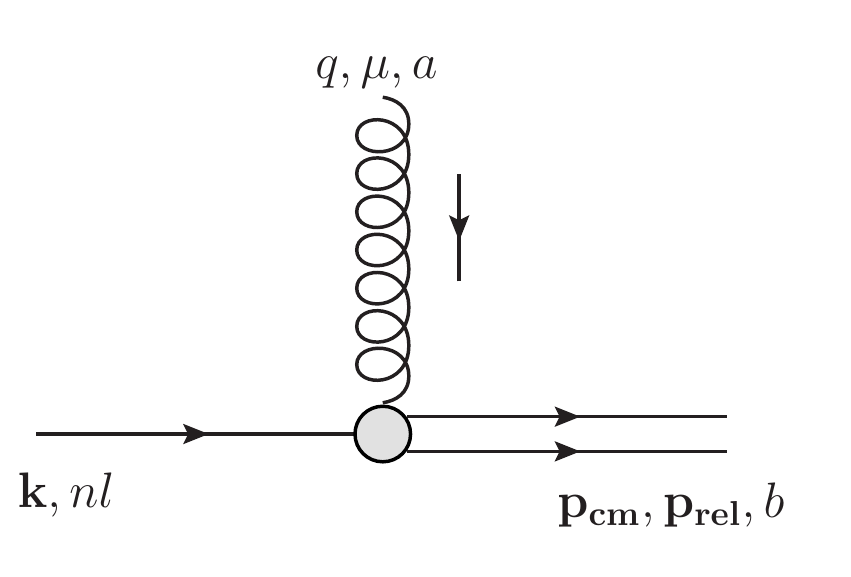}}    &=\ \ \ \ &  $g\sqrt{\frac{T_F}{N_c}}\delta^{ab}(q^0g^{\mu i}  - q^i g^{\mu0}) \langle \Psi_{{\bs p}_\ma{rel}} | r^i | \psi_{nl} \rangle$\\
	\raisebox{-0.26in}{  \includegraphics[height=1.3in]{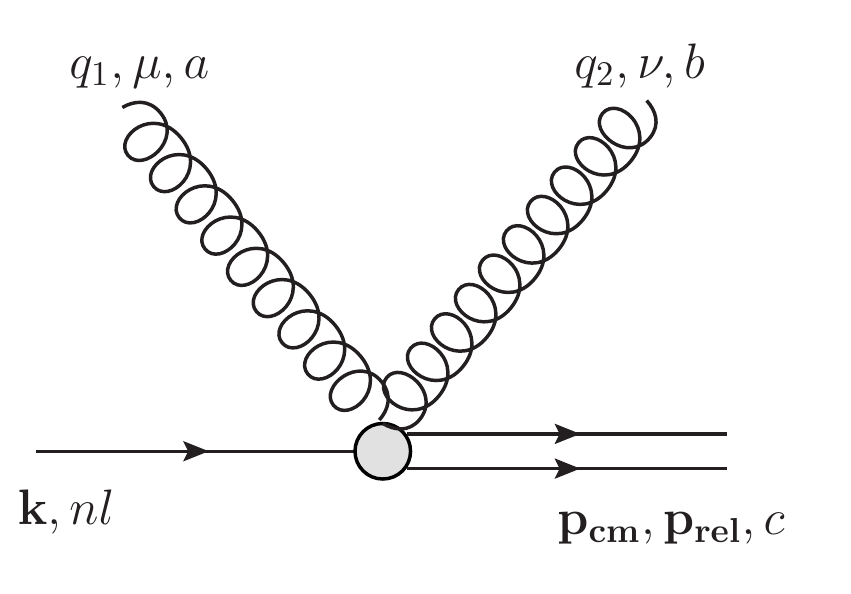}}   &=\ \ \ \ & $ig^2\sqrt{\frac{T_F}{N_c}}f^{abc}(g^{\mu 0}g^{\nu i}-g^{\mu i}g^{\nu 0}) \langle  \Psi_{{\bs p}_\ma{rel}} | r^i | \psi_{nl} \rangle $ \\
	 \raisebox{-0.26in}{ \includegraphics[height=1.3in]{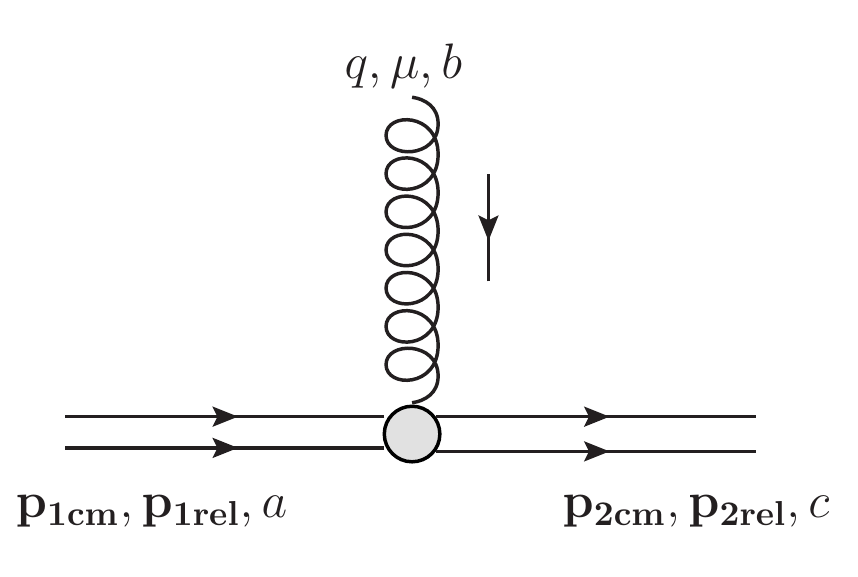} } &=\ \ \ \ &  $gd^{abc}(q^0g^{\mu i} - q^i g^{\mu0}) \langle  \Psi_{{\bs p}_\ma{2rel}}  | r^i | \Psi_{{\bs p}_\ma{1rel}} \rangle$ \\
	 \raisebox{-0.26in}{\includegraphics[height=1.3in]{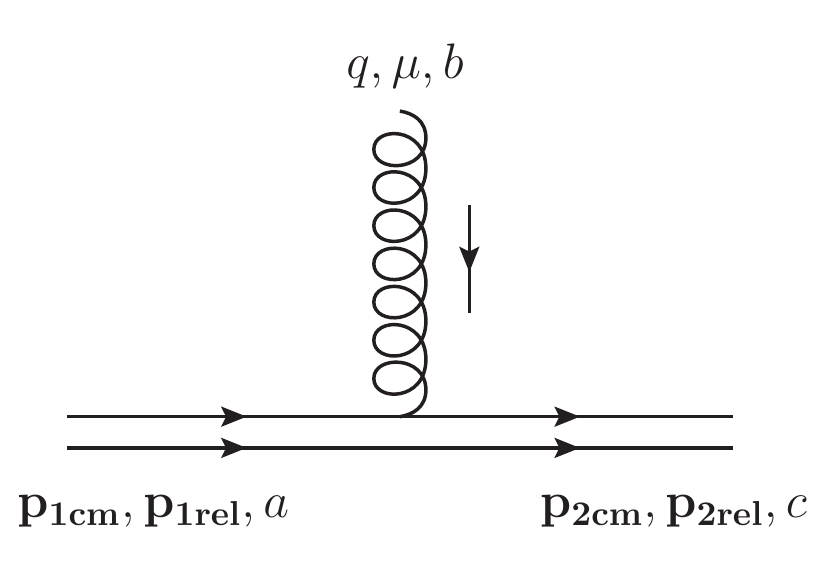} }   &=\ \ \ \ & $gf^{abc} g^{\mu0} (2\pi)^3\delta^3({\bs p}_\ma{rel1} - {\bs p}_\ma{rel2}) $
	\end{tabular}
\caption{Feynman rules in pNRQCD. Single solid line represents the bound color singlet while double solid lines represent the unbound color octet. The grey blob indicates the dipole interaction. The vertex with no grey blob means the gauge coupling in the c.m.~motion. Unbound singlet propagator will not be used throughout the paper and not shown here. The octet wavefunction $| \Psi_{{\bs p}_{\ma{rel}}} \rangle$ is a Coulomb scattering wave and thus the effect of the octet potential has been resummed in the octet propagator. In principle, there is also an octet-octet-gluon-gluon vertex at the order $g^2r$. But it is irrelevant in the current study and neglected here.}
\label{fig:rules}
\end{figure}

To make the wavefunction associated with the relative motion explicit, we do a change of basis in the relative motion by defining
\be
\ma{S}(\bs R, \bs r, t) &=& \frac{1}{\sqrt{N_c}}S(\bs R, \bs r, t) \equiv \frac{1}{\sqrt{N_c}}\langle \bs r | S(\bs R, t) \rangle \\
\ma{O}(\bs R, \bs r, t) &=& \frac{1}{\sqrt{T_F}}O^a(\bs R, \bs r, t)T^a  \equiv  \frac{1}{\sqrt{T_F}} \langle  \bs r | O^a(\bs R, t)\rangle T^a \,,
\ee
where $N_c=3$ and $T_F = \frac{1}{2}$. We define the quadratic Casimir of the fundamental representation $C_F \equiv \frac{T_F}{N_c}(N_c^2-1)$ for later use.
Then the Lagrangian density of the singlet and octet can be written as \cite{Fleming:2005pd}
\be
\label{eq:lagr2}
\ml{L}_\ma{pNRQCD}({\bs R}, t) &=&  \ml{L}_{\ma{kin},s} + \ml{L}_{\ma{kin},o} + \ml{L}_{\ma{int},so} + \ml{L}_{\ma{int},oo} +\cdots \\
\ml{L}_{\ma{kin},s}  &=&  \langle S(\bs R, t) | (i\partial_0-H_s) | S(\bs R, t)\rangle \\
\ml{L}_{\ma{kin},o}  &=&  \langle O^a(\bs R, t) | ( i\partial_0-H_o )  |O^a(\bs R, t)\rangle \\
\ml{L}_{\ma{int},so} &=&  \sqrt{\frac{T_F}{N_C}}\Big( \langle O^a(\bs R, t) | \bs r \cdot g{\bs E}^a(\bs R, t) | S(\bs R, t)\rangle + \ma{h.c.} \Big) \\
\ml{L}_{\ma{int},oo} &=& if^{abc} \langle O^a(\bs R, t) | g A_0^b(\bs R, t)   |O^c(\bs R, t)\rangle \\\nn
&+& d^{abc}  \langle O^{a}(\bs R, t)| g \bs r \cdot \bs E^b(\bs R, t) | O^{c}(\bs R, t)\rangle +\cdots  \, ,
\ee
The bra-ket notation saves us from writing the integral over the relative position explicitly. The singlet and octet composite fields are quantized by
\be \nn
|S(\bs R, t) \rangle &=& \int\frac{\diff^3 p_\ma{cm}}{(2\pi)^3}  e^{-i(Et-\bs p_\ma{cm} \cdot \bs R)} \bigg( \sum_{nl}  | \psi_{nl} \rangle a_{nl}(\bs p_\ma{cm})   + \int\frac{\diff^3 p_{\ma{rel}}}{(2\pi)^3}   | \psi_{{\bs p}_{\ma{rel}}} \rangle b_{{\bs p}_{\ma{rel}}}(\bs p_\ma{cm})  \bigg) \\
\\
|O^a(\bs R, t) \rangle &=&  \int\frac{\diff^3 p_\ma{cm}}{(2\pi)^3} e^{-i(Et-\bs p_\ma{cm}\cdot \bs R)}  \int\frac{\diff^3 p_{\ma{rel}}}{(2\pi)^3} 
| \Psi_{{\bs p}_{\ma{rel}}} \rangle c^a_{{\bs p}_{\ma{rel}}}(\bs p_\ma{cm})  \,,
\ee
where $E$ is the eigenenergy of the state under the Hamiltonians, Eq.~(\ref{eqn:hamiltonian}). The whole Hilbert space factorizes into two parts: one for the c.m.~motion and the other for the relative motion. 
The wavefunctions of the relative motion can be obtained by solving Schr\"odinger equations, which are part of the equations of motion of the free composite fields. 
They can be hydrogen-like wavefunctions $| \psi_{nl} \rangle$ for bound singlets with the eigenenergy $-|E_{nl}|$, or Coulomb scattering waves $| \psi_{{\bs p}_{\ma{rel}}} \rangle$ and $ | \Psi_{{\bs p}_{\ma{rel}}} \rangle $ for unbound singlets and octets with the eigenenergy ${\bs p}_\ma{rel}^2/M$. No bound state exists in the octet channel due to the repulsive potential. 
As we will explain later (see expression (\ref{eqn:average_m})), we will average over the third component of the angular momentum of non-$S$ wave quarkonium states. So we omit the quantum number $m_l$ of the bound singlet state.
The operators $a^{(\dagger)}_{nl}(\bs p_\ma{cm})$, $b^{(\dagger)}_{{\bs p}_{\ma{rel}}}(\bs p_\ma{cm})$ and $c^{a(\dagger)}_{{\bs p}_{\ma{rel}}}(\bs p_\ma{cm})$ act on the Fock space to annihilate (create) a composite particle with c.m.~momentum ${\bs p_\ma{cm}}$ and corresponding quantum numbers in the relative motion. 
These annihilation and creation operators satisfy the following commutation rules:
\be
[a_{n_1l_1}({\bs p}_{\ma{cm}1}),\ a^{\dagger}_{n_2l_2}({\bs p}_{\ma{cm}2})] &=& (2\pi)^3 \delta^3({\bs p}_{\ma{cm}1} - {\bs p}_{\ma{cm}2}) \delta_{n_1n_2}\delta_{l_1l_2} \\
{[b_{{\bs p}_{\ma{rel}1}}({\bs p}_{\ma{cm}1}),\ b^{\dagger}_{{\bs p}_{\ma{rel}2}}({\bs p}_{\ma{cm}2})]} & =& (2\pi)^6 \delta^3({\bs p}_{\ma{cm}1} - {\bs p}_{\ma{cm}2})\delta^3({\bs p}_{\ma{rel}1}-{\bs p}_{\ma{rel}2}) \\
{[c^{a_1}_{{\bs p}_{\ma{rel}1}}({\bs p}_{\ma{cm}1}),\ c^{a_2\dagger}_{{\bs p}_{\ma{rel}2}}({\bs p}_{\ma{cm}2})]} &=& (2\pi)^6 \delta^3({\bs p}_{\ma{cm}1} - {\bs p}_{\ma{cm}2}) \delta^3({\bs p}_{\ma{rel}1}-{\bs p}_{\ma{rel}2}) \delta^{a_1a_2}\,.
\ee
All other commutators are zero. The Feynman rules are summarized in Fig.~\ref{fig:rules}. We use the notation $a^ib^i \equiv \sum_i a^ib^i = a_ib_i$ to denote Euclidean summation and ${\bs r}^{\mu} = 0$ when $\mu=0$.

\section{Dissociation and recombination}
\label{sect:disso_reco}
In this section, we consider the contributions to the dissociation and recombination terms in the Boltzmann equation at the order $gr$ and $g^2r$. All contributing Feynman diagrams are shown in Fig.~\ref{fig:diagrams1}.

\begin{figure}
    \centering
    \begin{subfigure}[b]{0.33\textwidth}
        \centering
        \includegraphics[height=1.1in]{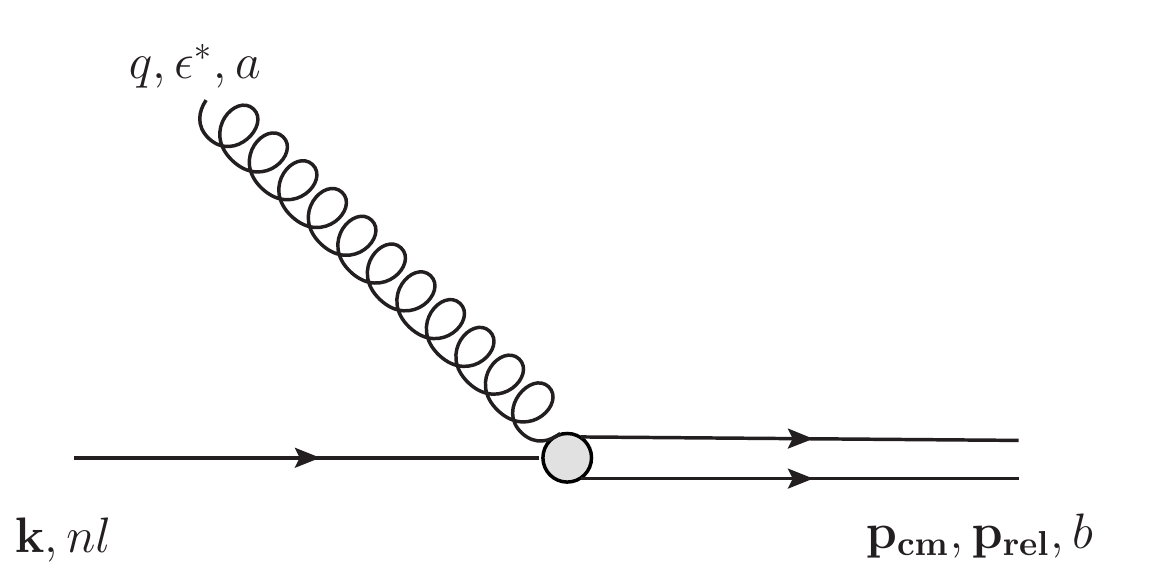}
        \caption{}\label{subfig:a}
    \end{subfigure}%
    ~ 
    \begin{subfigure}[b]{0.33\textwidth}
        \centering
        \includegraphics[height=1.25in]{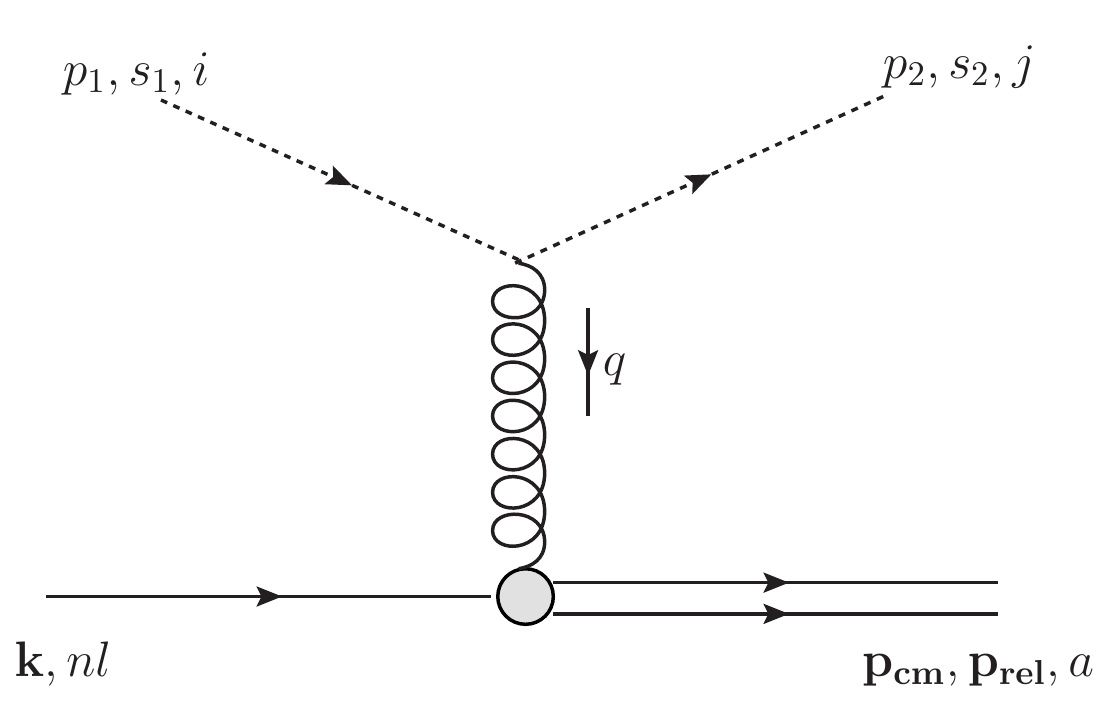}
        \caption{}\label{subfig:b}
    \end{subfigure}%
    ~ 
    \begin{subfigure}[b]{0.33\textwidth}
        \centering
        \includegraphics[height=1.25in]{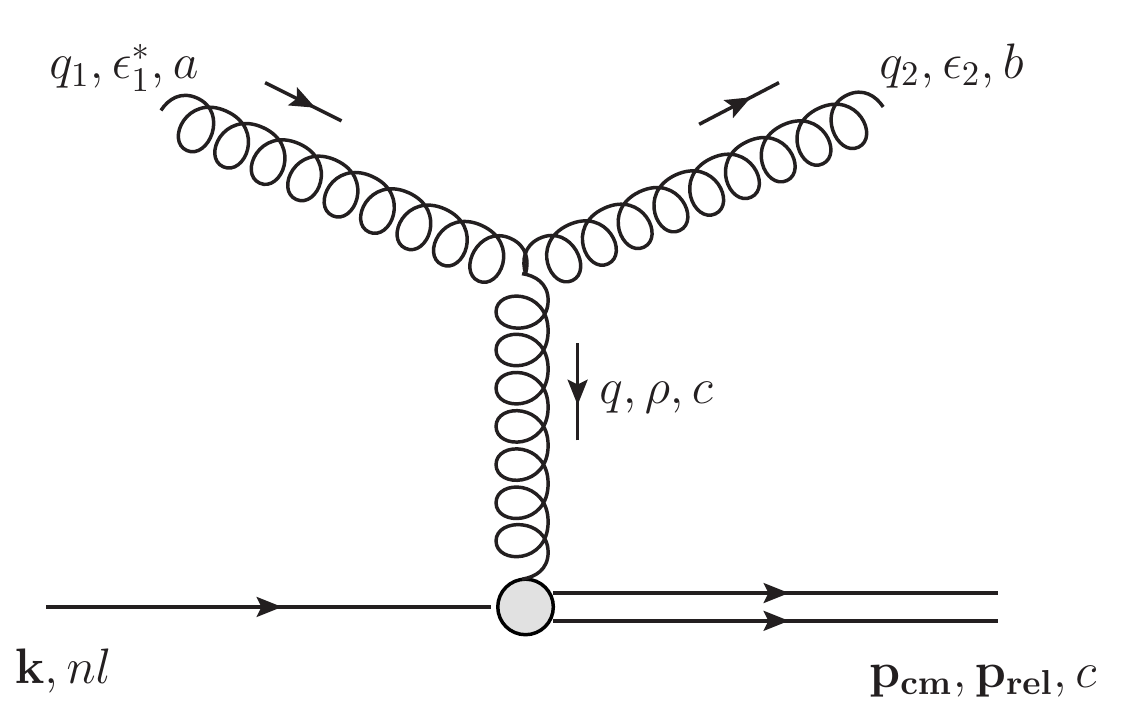}
        \caption{}\label{subfig:c}
    \end{subfigure}%

    \begin{subfigure}[b]{0.33\textwidth}
        \centering
        \includegraphics[height=1.25in]{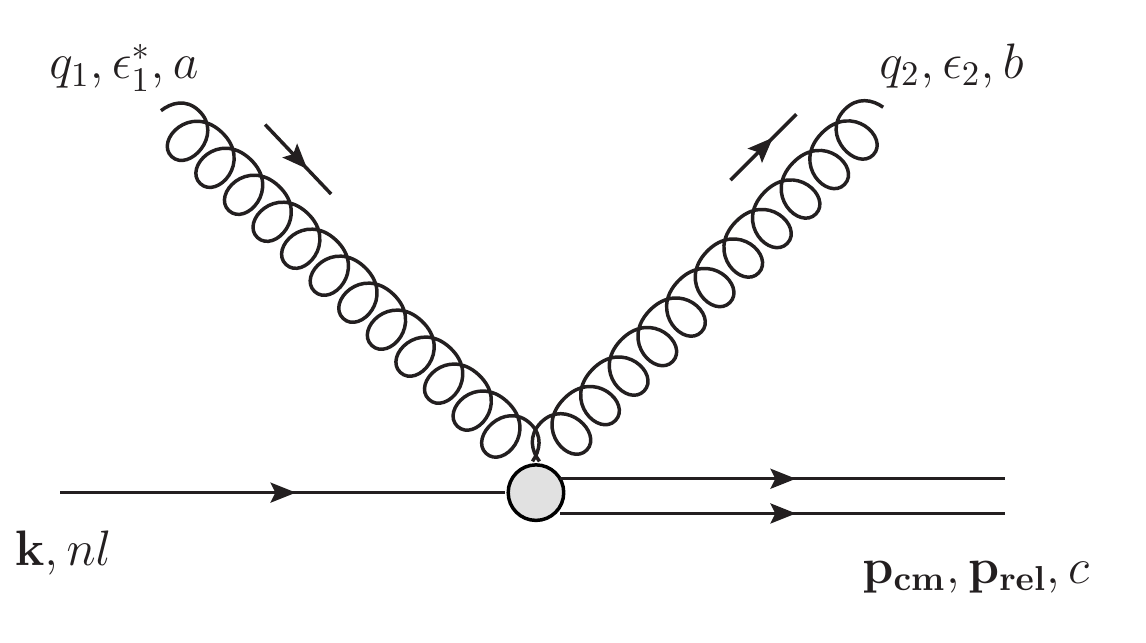}
        \caption{}\label{subfig:d}
    \end{subfigure}%
    ~ 
    \begin{subfigure}[b]{0.33\textwidth}
        \centering
        \includegraphics[height=1.25in]{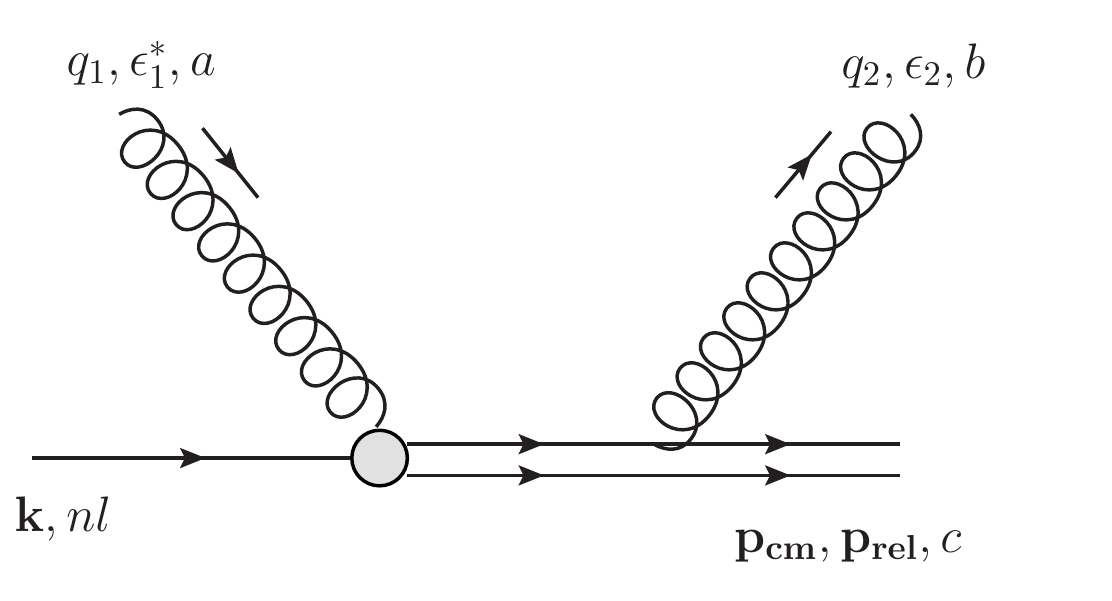}
        \caption{}\label{subfig:e}
    \end{subfigure}%
    ~ 
    \begin{subfigure}[b]{0.33\textwidth}
        \centering
        \includegraphics[height=1.25in]{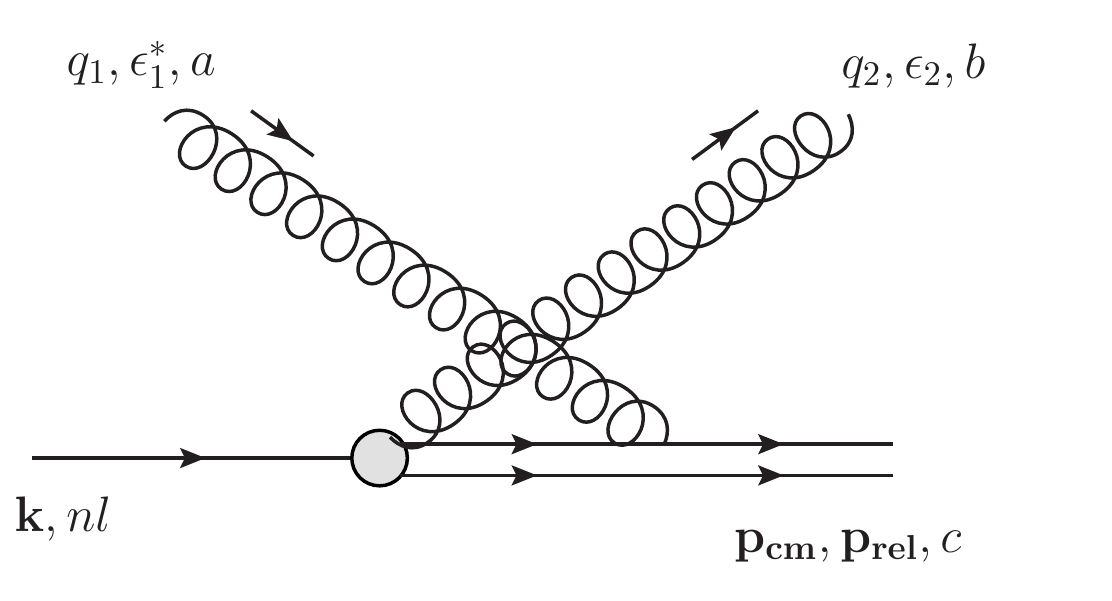}
        \caption{}\label{subfig:f}
    \end{subfigure}%
    
   \begin{subfigure}[b]{0.33\textwidth}
        \centering
        \includegraphics[height=1.25in]{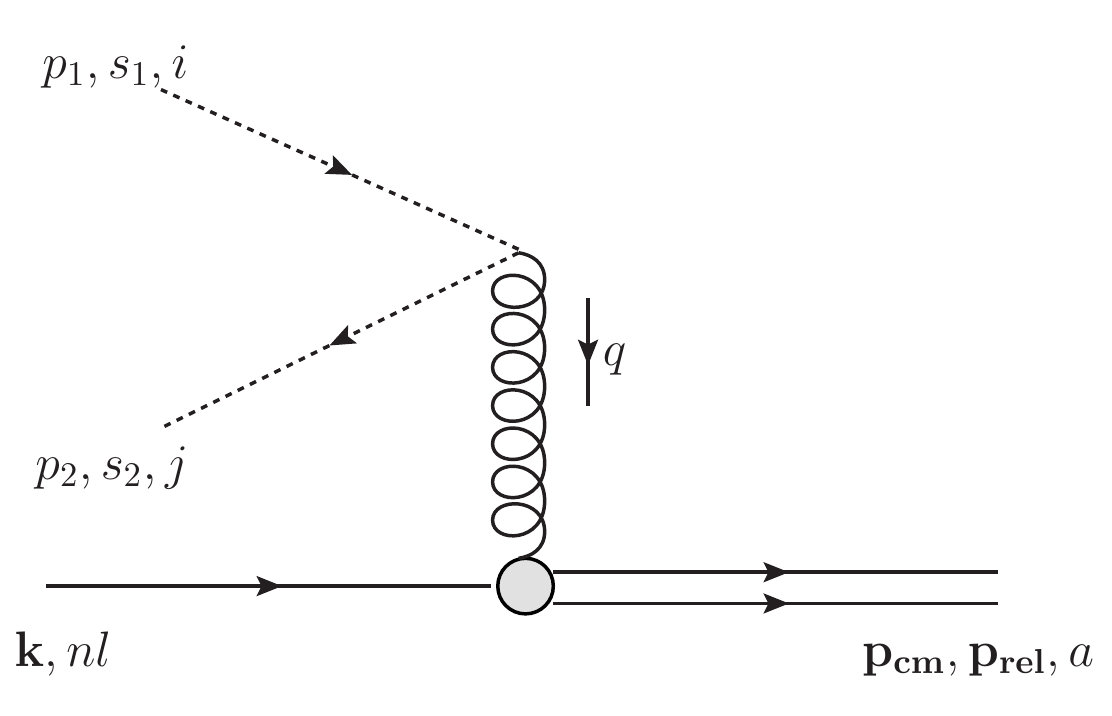}
        \caption{}\label{subfig:g}
    \end{subfigure}%
    ~ 
    \begin{subfigure}[b]{0.33\textwidth}
        \centering
        \includegraphics[height=1.25in]{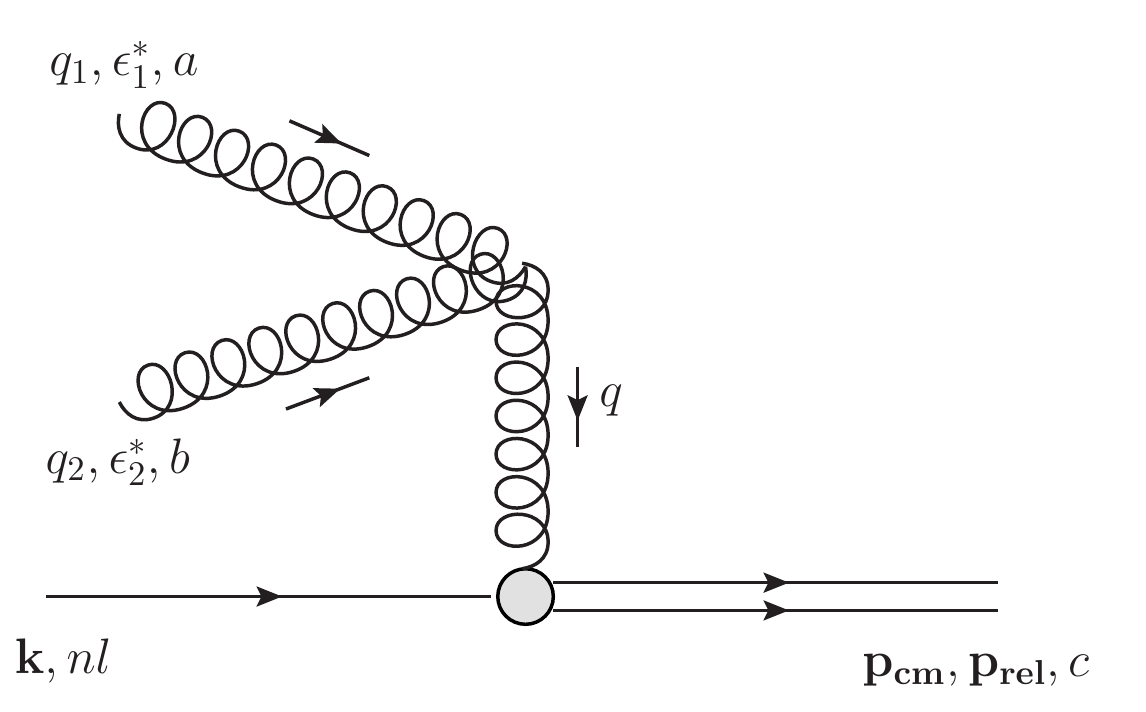}
        \caption{}\label{subfig:h}
    \end{subfigure}%
    ~  
    \begin{subfigure}[b]{0.33\textwidth}
        \centering
        \includegraphics[height=1.25in]{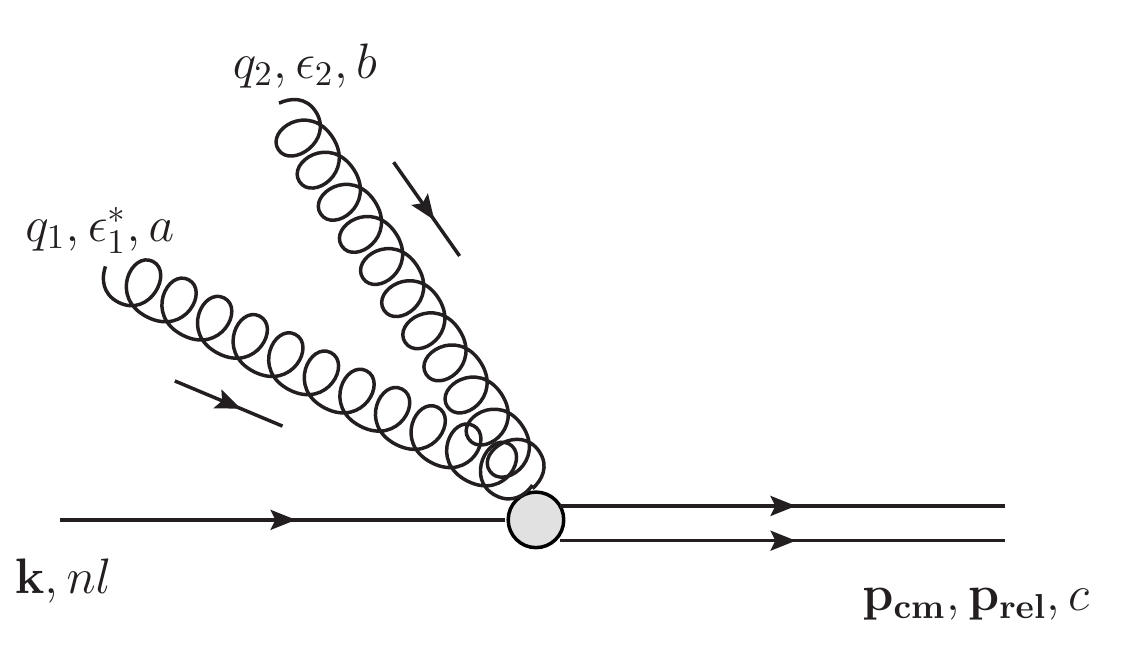}
        \caption{}\label{subfig:i}
    \end{subfigure}%
    
    \begin{subfigure}[b]{0.33\textwidth}
        \centering
        \includegraphics[height=1.25in]{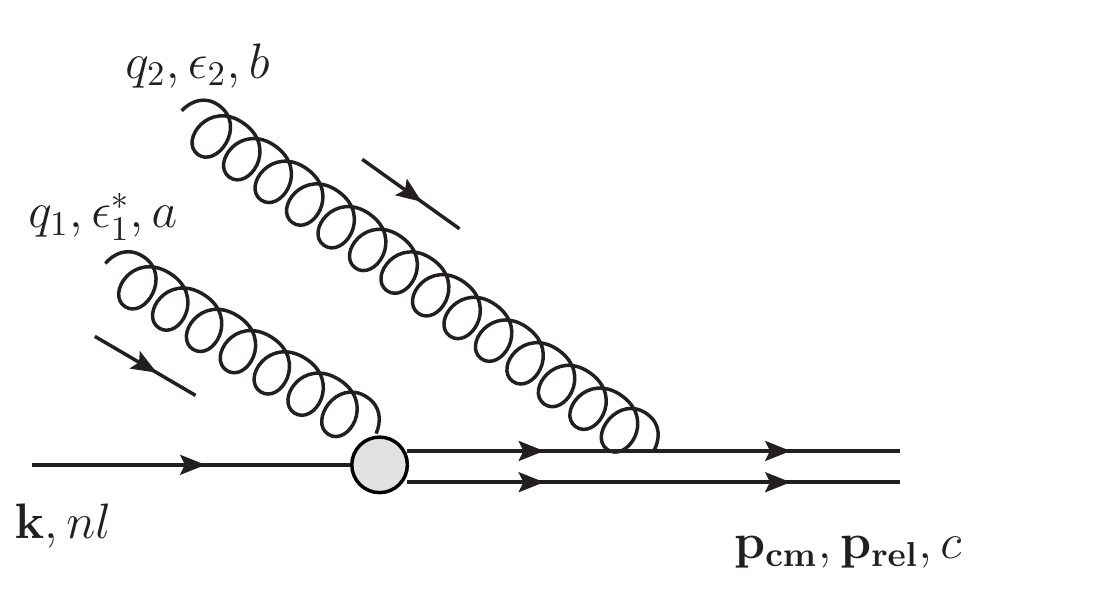}
        \caption{}\label{subfig:j}
    \end{subfigure}%
    ~
    \begin{subfigure}[b]{0.33\textwidth}
        \centering
        \includegraphics[height=1.25in]{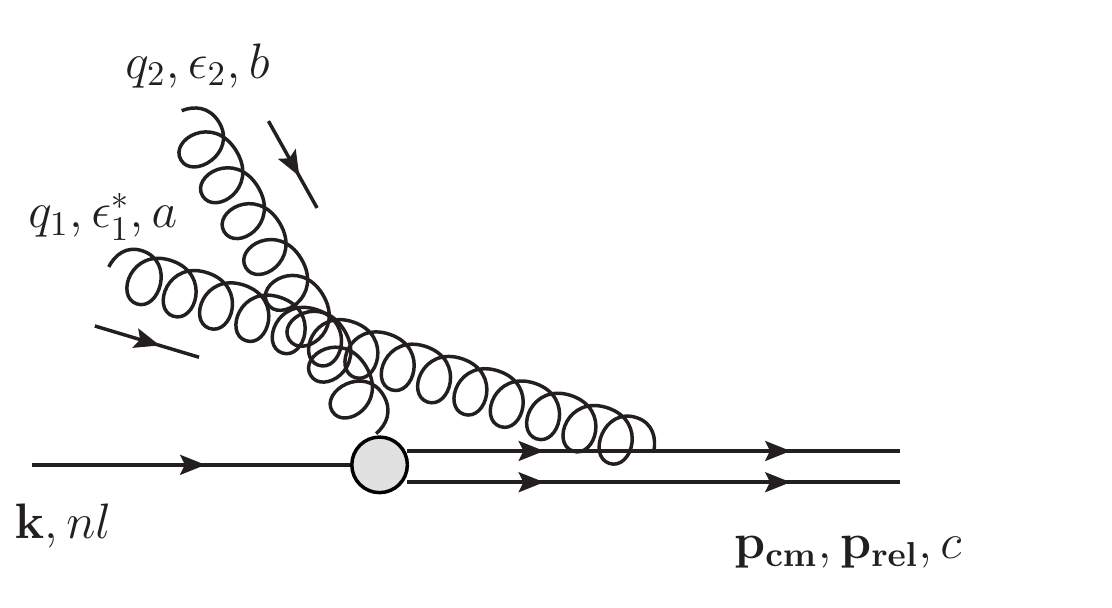}
        \caption{}\label{subfig:k}
    \end{subfigure}%
    ~      
   \begin{subfigure}[b]{0.33\textwidth}
        \centering
        \includegraphics[height=1.25in]{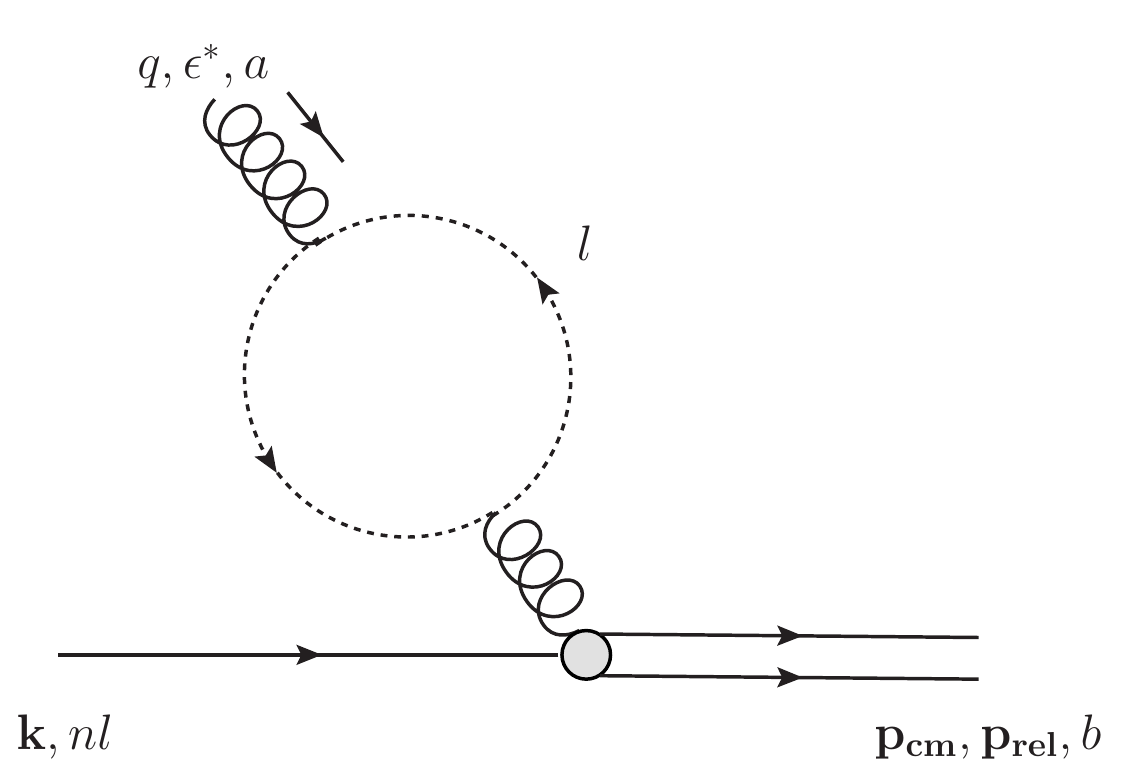}
        \caption{}\label{subfig:l}
    \end{subfigure}%
     
     \begin{subfigure}[b]{0.33\textwidth}
        \centering
        \includegraphics[height=1.25in]{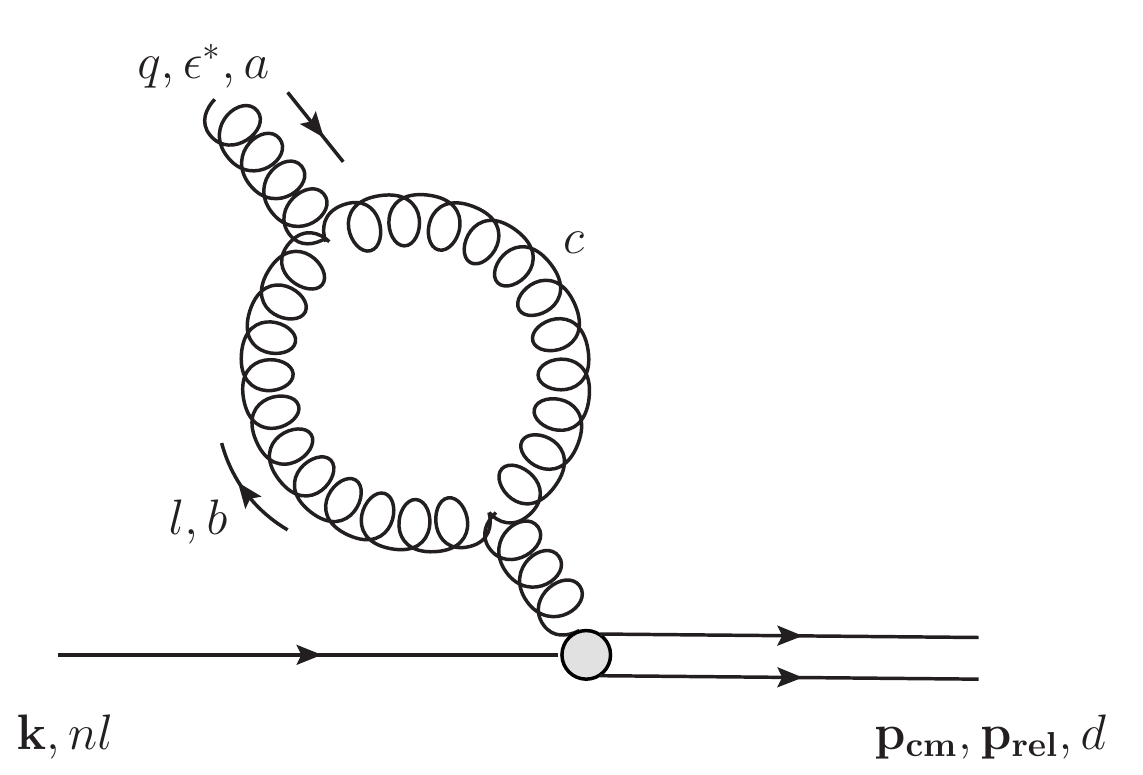}
        \caption{}\label{subfig:m}
    \end{subfigure}%
    ~ 
     \begin{subfigure}[b]{0.33\textwidth}
        \centering
        \includegraphics[height=1.25in]{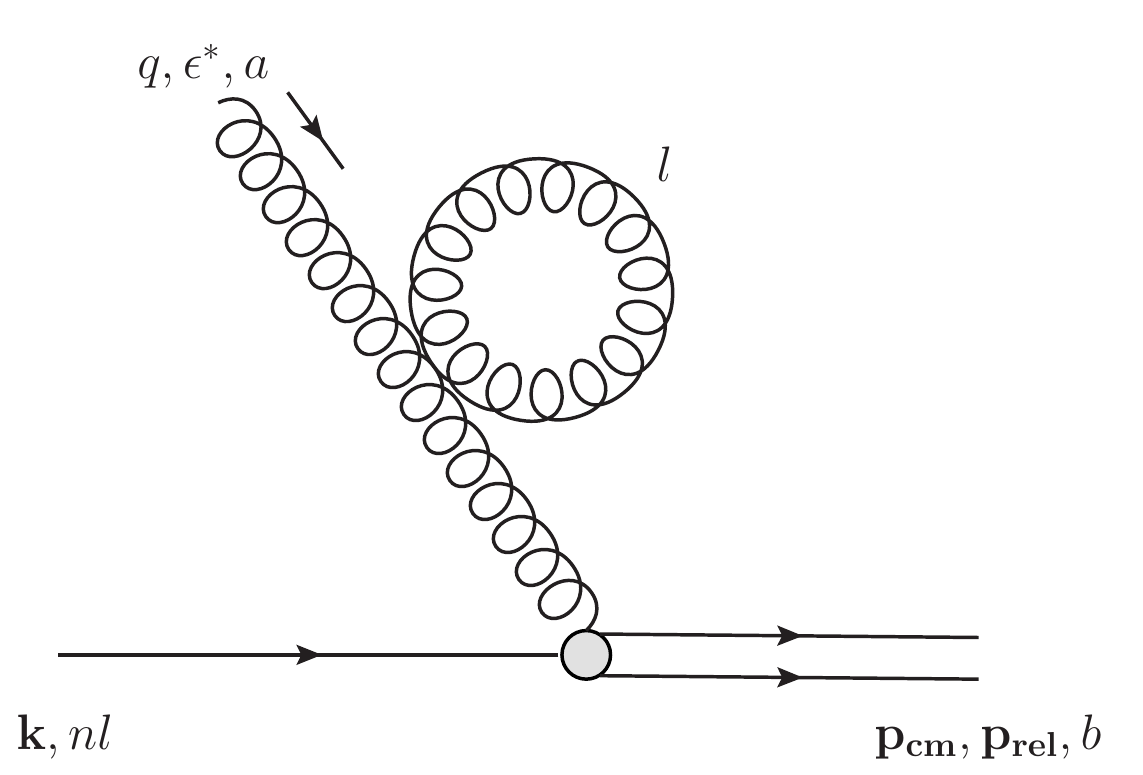}
        \caption{}\label{subfig:n}
    \end{subfigure}%
    ~
    \begin{subfigure}[b]{0.33\textwidth}
        \centering
        \includegraphics[height=1.25in]{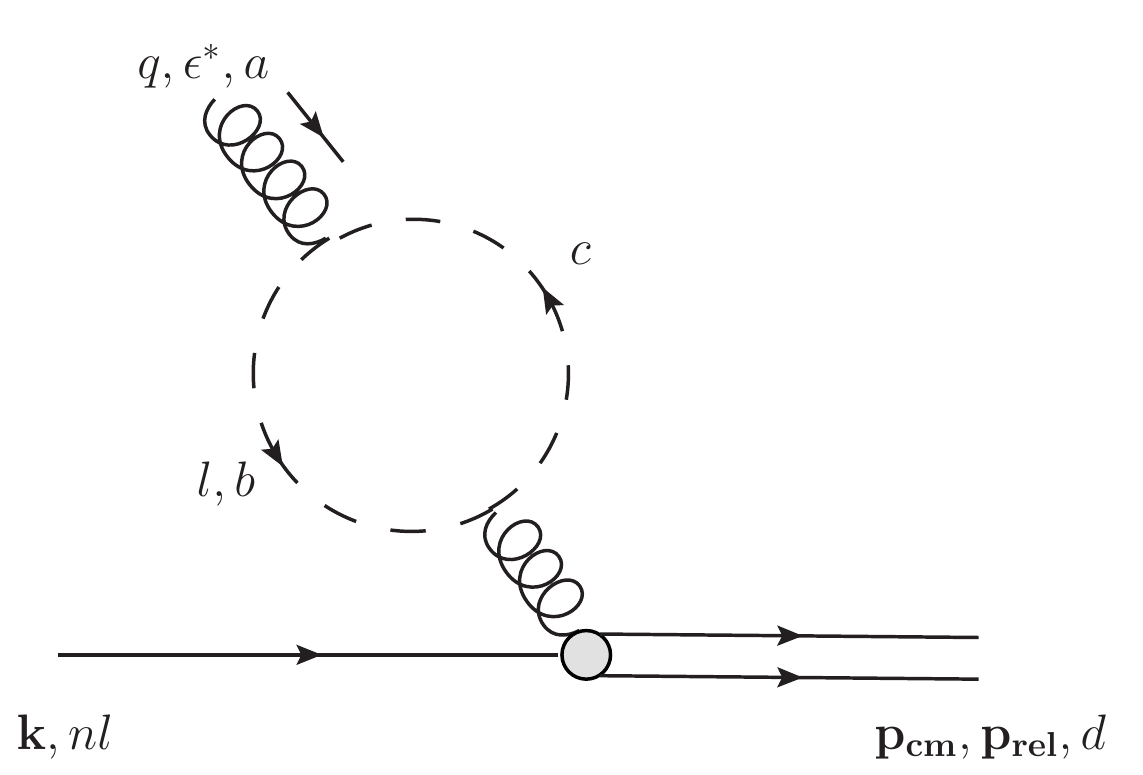}
        \caption{}\label{subfig:o}
    \end{subfigure}%
\end{figure}
\begin{figure}[htb]\ContinuedFloat        
     \begin{subfigure}[b]{0.33\textwidth}
        \centering
        \includegraphics[height=1.25in]{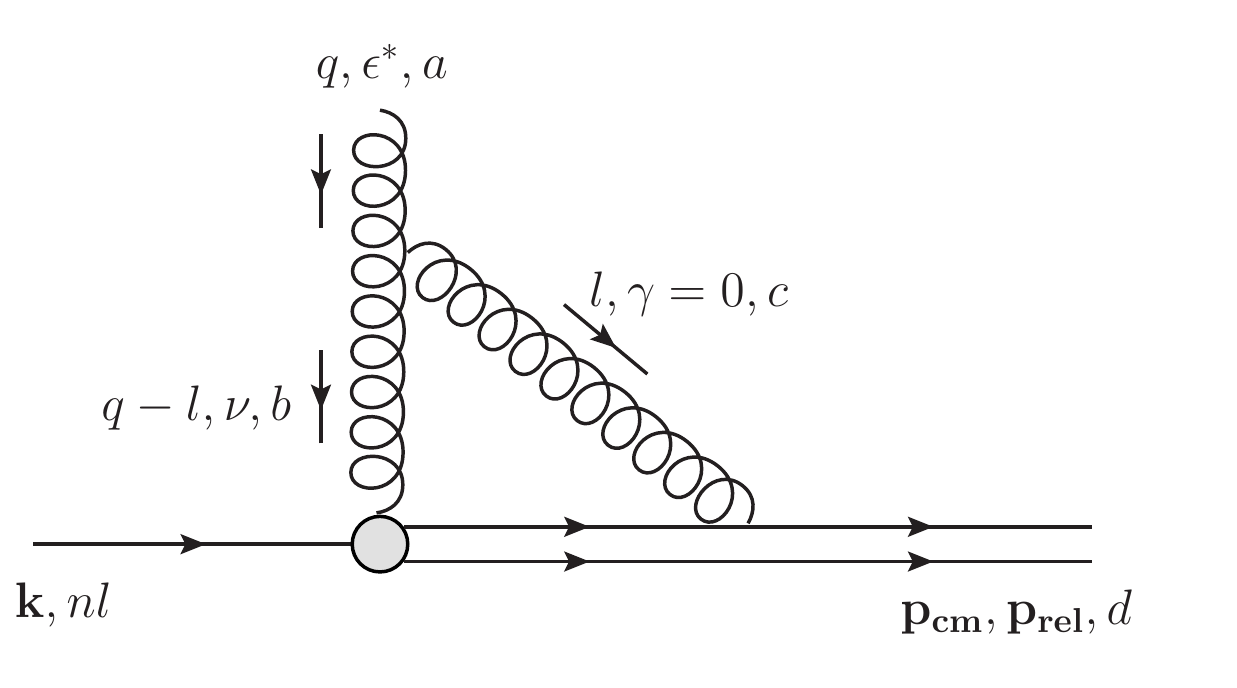}
        \caption{}\label{subfig:p}
    \end{subfigure}%
    ~   
     \begin{subfigure}[b]{0.33\textwidth}
        \centering
        \includegraphics[height=1.25in]{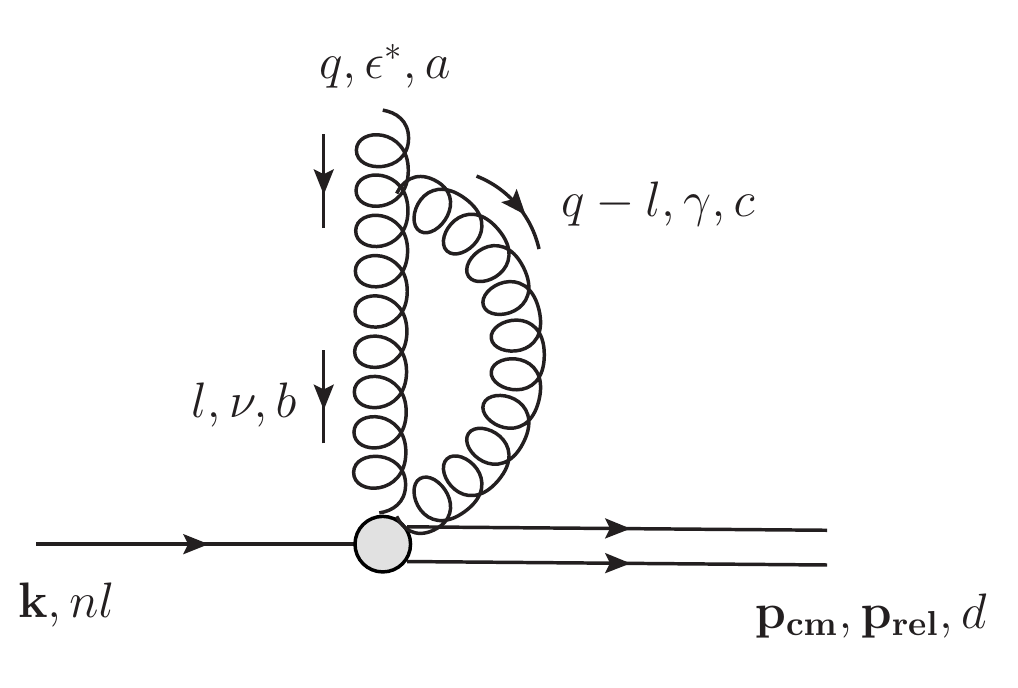}
        \caption{}\label{subfig:q}
    \end{subfigure}%
    ~    
    \begin{subfigure}[b]{0.33\textwidth}
        \centering
        \includegraphics[height=1.25in]{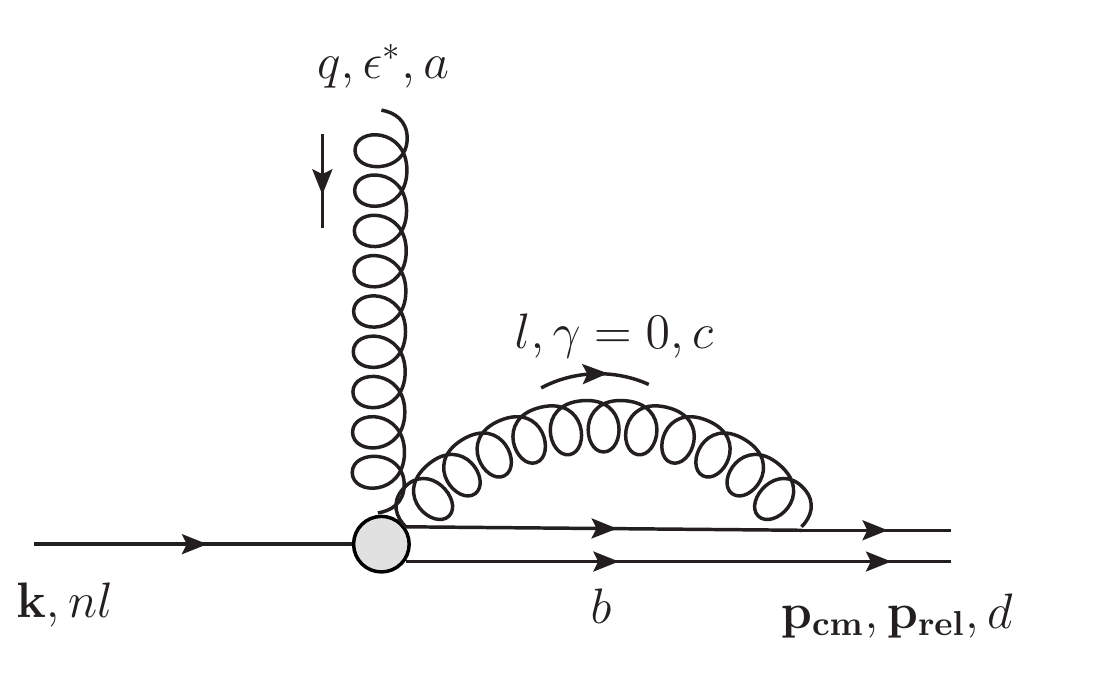}
        \caption{}\label{subfig:r}
    \end{subfigure}%

     \begin{subfigure}[b]{0.33\textwidth}
        \centering
        \includegraphics[height=1.25in]{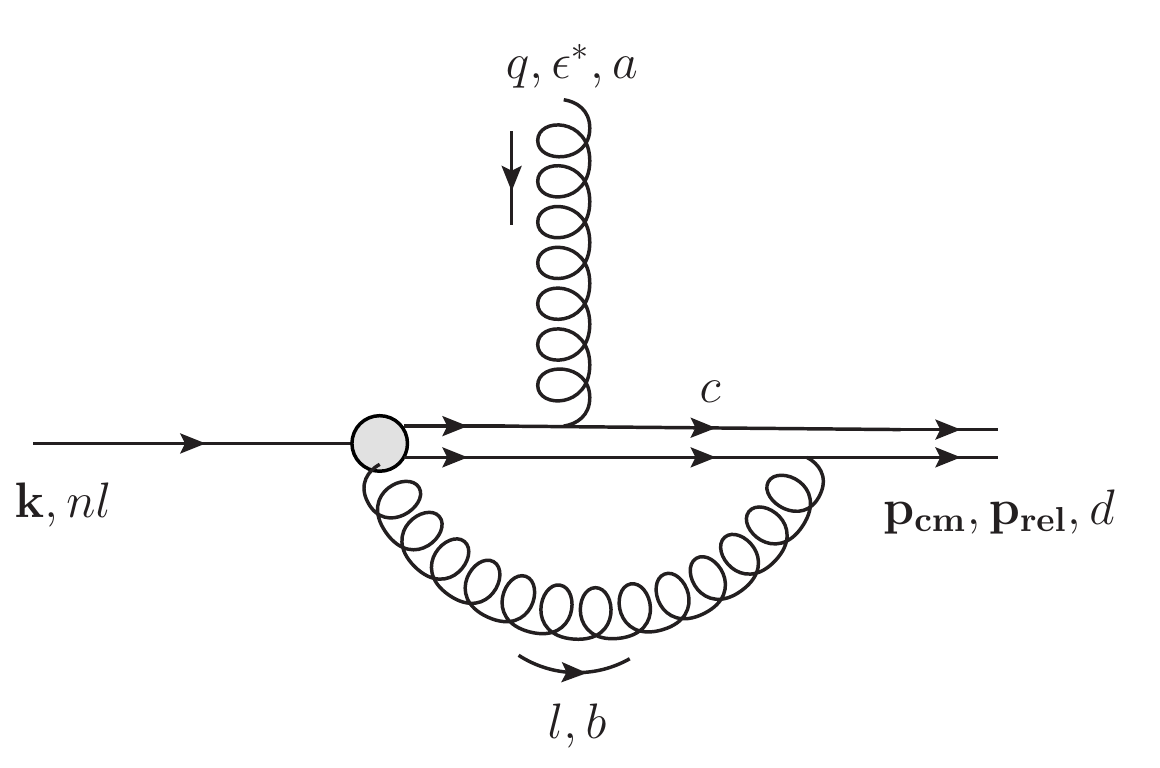}
        \caption{}\label{subfig:s}
    \end{subfigure}%
    ~
    \begin{subfigure}[b]{0.33\textwidth}
        \centering
        \includegraphics[height=1.25in]{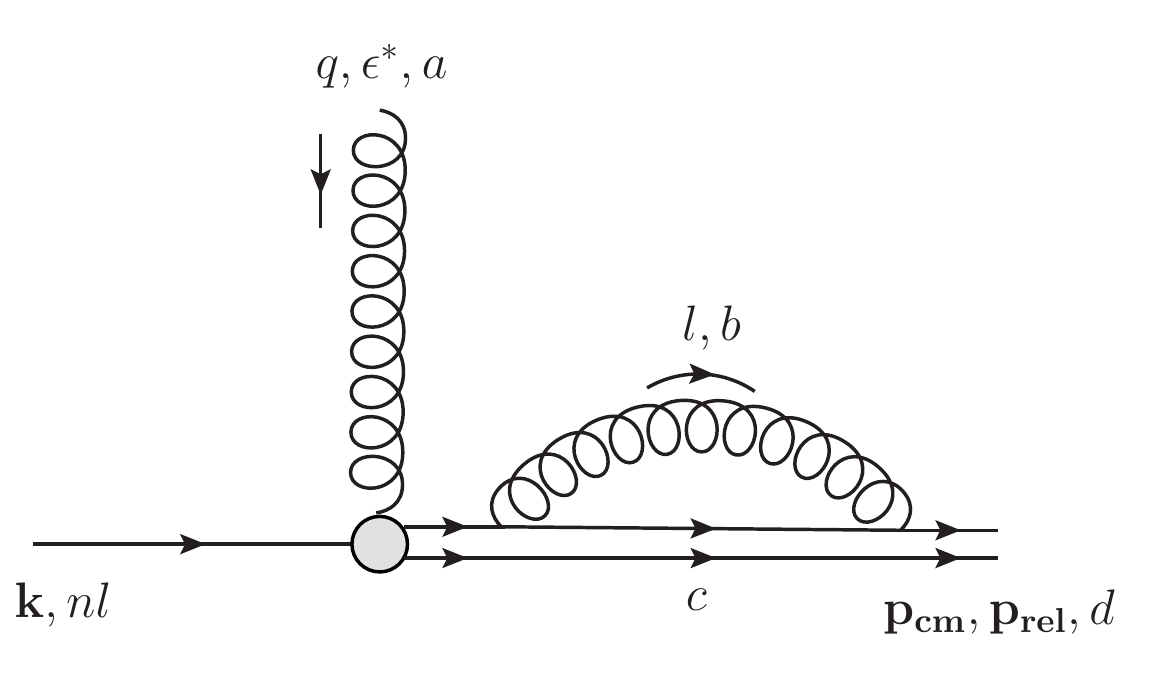}
        \caption{}\label{subfig:t}
    \end{subfigure}%
    \caption{Feynman diagrams contributing to the dissociation and recombination terms in the Boltzmann equation. Single solid line represents the bound color singlet while double solid lines represent the unbound color octet. Short dashed line indicates the light quark while the long dashed line is the ghost. The grey blob indicates the dipole interaction.}
    \label{fig:diagrams1}
\end{figure}

\subsection{Contributions at the order $gr$}
\label{subsect: a}
At the order $gr$, only the diagram in Fig.~\ref{subfig:a} contributes. It is a gluon absorption process for dissociation and a gluon emission process for recombination. The scattering amplitude is given by
\be
i\ml{M}_{(a)} &=& g\sqrt{\frac{T_F}{N_c}}(q^0\epsilon^{*i}  - q^i \epsilon^{*0}) \langle  \Psi_{{\bs p}_\ma{rel}}  | r^i |  \psi_{nl}  \rangle \delta^{ab}\\
& \equiv&  i \epsilon^{*\mu} (\ml{M}_{(a)})_\mu \,.
\ee
The gluon is on-shell so $q^0=|\bs q|\equiv q$. The Ward identity can be easily verified:
\be
\label{eqn:ward1}
 q^{\mu}  (\ml{M}_{(a)})_\mu= 0\,.
\ee
So we can compute the amplitude in any gauge we want. In Coulomb gauge, we define the square of the amplitude magnitude, summed over the gluon and octet colors $a,b$ and the gluon polarizations $\epsilon$
\be
\sum | \ml{M}_{(a)} |^2 \equiv \sum_{a,b,\epsilon}  | \ml{M}_{(a)} |^2 = g^2C_Fq^2(\delta^{ij}-\hat{q}^i\hat{q}^j)  \langle \psi_{nl} | r^i | \Psi_{{\bs p}_\ma{rel}} \rangle  \langle \Psi_{{\bs p}_\ma{rel}} | r^j |  \psi_{nl}  \rangle \,.
\ee
As mentioned earlier, we average over the third component of angular momentum of quarkonium for $l>0$ so we omit the quantum number $m_l$. Here we will show explicitly how the average simplifies the calculation. To this end, we temporarily restore the quantum number $m_l$ in the bound state wavefunction. When integrating over the relative momentum of the heavy quark antiquark pair from the dissociation, the average leads to
\be
\nn
&&\frac{1}{2l+1}\sum_{m_l=-l}^l \int \diff^3 p_{\ma{rel}} \langle \psi_{nlm_l} | r^i | \Psi_{{\bs p}_\ma{rel}} \rangle  \langle \Psi_{{\bs p}_\ma{rel}} | r^j |  \psi_{nlm_l}  \rangle \\ 
\label{eqn:average_m} 
&=& \frac{1}{3}\delta^{ij}  \frac{1}{2l+1}\sum_{m_l=-l}^l \int \diff^3 p_{\ma{rel}}    |  \langle \Psi_{{\bs p}_\ma{rel}}  | {\bs r} |   \psi_{nlm_l}  \rangle | ^2
\equiv \frac{1}{3}\delta^{ij}  \int \diff^3 p_{\ma{rel}} | \langle \Psi_{{\bs p}_\ma{rel}}  | {\bs r} |   \psi_{nl} \rangle |^2\,.
\ee
This allows us to write
\be
\sum | \ml{M}_{(a)} |^2 \equiv  \frac{2}{3}g^2C_Fq^2 |  \langle   \Psi_{{\bs p}_\ma{rel}} | {\bs r} |   \psi_{nl}  \rangle |^2\,.
\ee

To simplify the notation of the dissociation and recombination terms for a quarkonium state $nls$ in the Boltzmann equation, we define\footnote{In $\ml{F}^+_{nls(a)}$, the positions in the heavy quark and antiquark distributions can be different. See the derivation of this term in Ref.~\cite{Yao:2018nmy}.}
\be\nn
\ml{F}^+_{nls(a)} & \equiv & g_+ \int \frac{\diff^3 k}{(2\pi)^3}  \frac{\diff^3 p_{Q}}{(2\pi)^3}  \frac{\diff^3 p_{\bar{Q}} }{(2\pi)^3} \frac{\diff^3 q}{2q(2\pi)^3} (1+n_B(q)) f_Q({\bs x}_Q, {\bs p}_Q,t) f_{\bar{Q}} ({\bs x}_{\bar{Q}}, {\bs p}_{\bar{Q}},t) \\
&& (2\pi)^4\delta^3({\bs k} + {\bs q} - {\bs p}_{\ma{cm}}) \delta(-|E_{nl}|+q-\frac{p^2_{\ma{rel}}}{M})  \sum | \ml{M}_{(a)} |^2\\ \nn
\ml{F}^-_{nls(a)} & \equiv &  \int \frac{\diff^3 k}{(2\pi)^3}  \frac{\diff^3 p_{\ma{cm}}}{(2\pi)^3}  \frac{\diff^3 p_{\ma{rel}}}{(2\pi)^3} \frac{\diff^3 q}{2q(2\pi)^3} n_B(q) f_{nls}({\bs x}, {\bs k}, t) \\
&& (2\pi)^4\delta^3({\bs k} + {\bs q} - {\bs p}_{\ma{cm}}) \delta(-|E_{nl}|+q-\frac{p^2_{\ma{rel}}}{M})   \sum | \ml{M}_{(a)} |^2\,,
\ee
where $n_B$ is the Bose-Einstein distribution function, $g_+ = \frac{1}{N_c^2}g_s$ and $g_s$ is the multiplicity factor in spin: $g_s = \frac{3}{4}$ for a quarkonium with spin $s=1$ and $\frac{1}{4}$ for spin $s=0$. In the definition of $\ml{F}^+_{nls(a)}$, ${\bs p}_{\ma{cm}} = {\bs p}_Q+{\bs p}_{\bar{Q}},$ and ${\bs p}_{\ma{rel}} = \frac{{\bs p}_Q - {\bs p}_{\bar{Q}}}{2}$ are the c.m.~and relative momenta of a pair of heavy quark and antiquark with momenta ${\bs p}_Q$ and 
${\bs p}_{\bar{Q}}$.

We further define a ``$\delta-$derivative" symbol, first introduced in Ref.~\cite{Yao:2018zze}
\be
\frac{\delta }{\delta{{\bs p}_i}} \int \prod_{j=1}^n \frac{\Diff{3}p_j}{(2\pi)^3} h({\bs p}_1, {\bs p}_2, \cdots, {\bs p}_n)\Big|_{{\bs p}_i = {\bs p}} 
&\equiv&  \frac{\delta }{\delta{w({\bs p}})} \int \prod_{j=1}^n \frac{\Diff{3}p_j}{(2\pi)^3} h({\bs p}_1, {\bs p}_2, \cdots, {\bs p}_n)w({\bs p}_i) \\ \nn
&=& \int \prod_{j=1, j\neq i}^n \frac{\Diff{3}p_j}{(2\pi)^3} h({\bs p}_1, {\bs p}_2, \cdots, {\bs p}_{i-1}, {\bs p}, {\bs p}_{i+1}, \cdots, {\bs p}_n)\,,
\ee
where the second $\delta$ denotes the standard functional variation and $h({\bs p}_1, {\bs p}_2, \cdots, {\bs p}_n)$ and $w({\bs p}_i)$ are arbitrary independent smooth functions. Then the $\ml{C}^{\pm}_{nls}$ terms in the Boltzmann equation (\ref{eqn:boltzmann}) can be written as
\be
\ml{C}_{nls(a)}^{\pm}({\bs x}, {\bs p}, t) &=&  \frac{\delta \ml{F}^\pm_{nls(a)}}{\delta{{\bs k}}}  \Big|_{{\bs k}={\bs p}}\,.
\ee
For $\ml{C}_{nls(a)}^{+}$, we further require $\frac{1}{2}({\bs x}_Q + {\bs x}_{\bar{Q}}) = {\bs x}$, i.e., the position of a recombined quarkonium is given by the c.m.~position of the heavy quark antiquark pair. Recombinations from the heavy quark antiquark pairs with a distance much larger than the Bohr radius $|{\bs x}_Q - {\bs x}_{\bar{Q}} | \gg a_B$ can be shown to be negligible \cite{Yao:2018nmy}. In practical numerical simulations, a Gaussian dependence on $\frac{|{\bs x}_Q - {\bs x}_{\bar{Q}}|}{  a_B}$ may be applied \cite{Yao:2017fuc}. 

The dissociation rate of the quarkonium state $nls$ from the diagram in Fig.~\ref{subfig:a} is given by
\be
\Gamma_{nls(a)}^{\ma{disso}}({\bs x}, {\bs p}, t) \equiv \frac{\ml{C}_{nls(a)}^{-}({\bs x}, {\bs p}, t)}{ f_{nls}({\bs x}, {\bs p}, t) }\,.
\ee
The recombination rate of a heavy quark into the quarkonium state $nls$ surrounded by heavy antiquarks with the distribution $f_{\bar{Q}} ({\bs x}_{\bar{Q}}, {\bs p}_{\bar{Q}},t)$ is given by
\be
\Gamma_{nls(a)}^{\ma{recom}}({\bs x}, {\bs p}, t) \equiv \frac{1}{f_Q({\bs x}_Q, {\bs p}_Q,t) }  \frac{\delta \ml{F}^+_{nls(a)}}{\delta{{\bs p}_Q}} \Big|_{{\bs x}_Q={\bs x},\ {\bs p}_Q={\bs p}}\,.
\ee

\subsection{Contributions at the order $g^2r$}

\subsubsection{Contributions from diagram \ref{subfig:b}}
\label{subsect: b}
Fig.~\ref{subfig:b} depicts the inelastic scattering with light quarks (up and down) in the medium. The light quarks are assumed massless. First we check the amplitude is independent of the gauge choice. The gauge invariance reflects in the invariance of the amplitude when the $\epsilon_\mu$ in the gluon propagator is replaced with $\epsilon_\mu + q_\mu$. It has been shown in the last section \ref{subsect: a} that the dipole interaction between the singlet and octet is invariant under such a replacement, by virtue of the Ward identity Eq.~(\ref{eqn:ward1}). What remains to be shown is the invariance of the vertex of the light quark. This is guaranteed by the Dirac equation:
\be
\bar{u}_{s_2}(p_2)\gamma^\mu T^a u_{s_1}(p_1) q_\mu = \bar{u}_{s_2}(p_2)(\slashed{p}_1 - \slashed{p}_2 )T^au_{s_1}(p_1) = 0\,.
\ee
In Coulomb gauge,
\be \nn
i\ml{M}_{(b)} &=&  g^2V_A \sqrt{\frac{T_F}{N_c}} \langle \Psi_{{\bs p}_\ma{rel}} | r^k |  \psi_{nl}  \rangle
 \Big[ \frac{-q^0 (\delta^{kl} - \hat{q}^k\hat{q}^l) }{(q^0)^2-{\bs q}^2+i\epsilon}  \bar{u}_{s_2}(p_2) \gamma^l T^a u_{s_1}(p_1)
+ \frac{q^k}{ {\bs q}^2} \bar{u}_{s_2}(p_2)\gamma^0 T^a u_{s_1}(p_1)   \Big]\,. \\
&&
\ee
We define $|\ml{M}_{(b)}|^2$ summed over the octet color $a$, the spins $s_1,s_2$, colors $i,j$ and flavors of the incoming and outgoing light quarks and include also the contribution from antiquarks as
\be
\label{eqn:collinear_q}
&& \sum |\ml{M}_{(b)}|^2 \equiv \sum_{a,i,j} \sum_{s_1,s_2} \sum_{u,\bar{u},d,\bar{d}} |\ml{M}_{(b)}|^2 \\ 
&=& \frac{16}{3}g^4V_A^2T_FC_F  |\langle \Psi_{{\bs p}_\ma{rel}} | {\bs r} |  \psi_{nl}  \rangle|^2 \Big[  \frac{p_1p_2 + {\bs p}_1\cdot {\bs p}_2}{{\bs q}^2} 
 + \frac{2(q^0)^2(p_1p_2-{\bs p}_1\cdot\hat{q} \cdot {\bs p}_2\cdot \hat{q})}{((q^0)^2-{\bs q}^2+i\epsilon)^2} \Big] \,.
\ee

Next we check the infrared sensitivity of the term inside the square bracket. The energy-momentum conservation gives $q^0 = p_1 - p_2 = |{\bs p}_1| - |{\bs p}_2| = |E_{nl}| + p^2_{\ma{rel}}/M$, ${\bs q} = {\bs p}_1 - {\bs p}_2 = {\bs p}_{\ma{cm}}-{\bs k}$. Assume the angle between ${\bs p}_1$ and ${\bs p}_2$ is $\theta$. We have
\be
\frac{p_1p_2 + {\bs p}_1\cdot {\bs p}_2}{{\bs q}^2} &=& \frac{p_1p_2+p_1p_2\cos\theta}{p_1^2+p_2^2-2p_1p_2\cos\theta} \\
\frac{2(q^0)^2(p_1p_2-{\bs p}_1\cdot\hat{q} \cdot {\bs p}_2\cdot \hat{q})}{((q^0)^2-{\bs q}^2+i\epsilon)^2}  &=& \frac{p_1^2+p_2^2-2p_1p_2}{2p_1p_2(1-\cos\theta)}\times
\frac{p_1^2+p_2^2+p_1p_2(1-\cos\theta)}{p_1^2+p_2^2-2p_1p_2\cos\theta}\,.
\ee
In both terms, there is no soft divergence because the binding energy $|E_{nl}|$ serves as a soft regulator: $p_1^2+p_2^2-2p_1p_2\cos\theta \geq (p_1-p_2)^2 \geq |E_{nl}|^2$. The first term has no collinear divergence either. The collinear divergence happens in the second term when $\cos\theta\rightarrow1$. Physically this occurs when the momenta of both the incoming and outgoing light quarks are in the same direction. The transferred gluon is on-shell. In this case, the inelastic scattering cannot be distinguished from the real gluon process shown in Fig.~\ref{subfig:a}. As we show below in Section~\ref{subsect: lmno}, the interference between the diagram in Fig.~\ref{subfig:a} and its thermal loop correction Fig.~\ref{subfig:l} cancels this collinear divergence.

\subsubsection{Contributions from diagrams \ref{subfig:c}, \ref{subfig:d}, \ref{subfig:e} and \ref{subfig:f}}
\label{subsect: cdef}
The processes of inelastic scattering with gluons in the medium are depicted in Figs.~\ref{subfig:c}, \ref{subfig:d}, \ref{subfig:e} and \ref{subfig:f}. All the four diagrams are needed for the gauge invariance. First we consider the gauge transformation of the internal gluon line in Fig.~\ref{subfig:c}. If we cut the diagram into two halves by cutting the internal gluon line, we need to show both the upper and lower parts vanish when contracted with $q_\rho$. For the dipole interaction in the lower part, this has been shown by the Ward identity Eq.~(\ref{eqn:ward1}). For the three-gluon vertex in the upper part, it can be shown that
\be
-gf^{abc}(\epsilon^*_1)_\mu (\epsilon_2)_\nu\Big[  g^{\nu\rho}(q_2-q)^\mu + g^{\rho\mu}(q+q_1)^\nu + g^{\mu\nu}(-q_1-q_2)^\rho  \Big] q_{\rho} = 0\,,
\ee
by using $q_1 \cdot \epsilon_1 = q_2 \cdot \epsilon_2 =0$.

Next we consider the gauge transformation of the external gluon line. We fix the internal gluon line to be in the Lorentz gauge.
\be \nn
i\ml{M}_{(c)} \equiv i\ml{M}^{\mu\nu}_{(c)} (\epsilon_1^*)_\mu (\epsilon_2)_\nu
&=&-g^2V_A\sqrt{\frac{T_F}{N_c}} f^{abc} \Big[  g^{\nu\rho}(q_2-q)^\mu + g^{\rho\mu}(q+q_1)^\nu + g^{\mu\nu}(-q_1-q_2)^\rho  \Big]  \\
&&\frac{-ig_{\rho\sigma}}{(q_0)^2-{\bs q}^2+i\epsilon} (q^0\delta^{\sigma i} - q^i \delta^{\sigma0})  \langle \Psi_{{\bs p}_\ma{rel}} | r^i |  \psi_{nl}  \rangle (\epsilon_1^*)_\mu (\epsilon_2)_\nu 
\\
i\ml{M}_{(d)} \equiv i\ml{M}^{\mu\nu}_{(d)} (\epsilon_1^*)_\mu (\epsilon_2)_\nu &=& ig^2V_A\sqrt{\frac{T_F}{N_c}} f^{abc}\Big[ (\epsilon_1^*)^0 (\epsilon_2)^i - (\epsilon_1^*)^i (\epsilon_2)^0 \Big] \langle \Psi_{{\bs p}_\ma{rel}} | r^i |  \psi_{nl}  \rangle
\\ \nn
i\ml{M}_{(e)} \equiv i\ml{M}^{\mu\nu}_{(e)} (\epsilon_1^*)_\mu (\epsilon_2)_\nu &=& g^2V_A\sqrt{\frac{T_F}{N_c}} f^{abc}(\epsilon_2)^0 \Big[ (q_1)^0 (\epsilon_1^*)^i - (q_1)^i (\epsilon_1^*)^0 \Big]
\langle \Psi_{{\bs p}_\ma{rel}} | r^i |  \psi_{nl}  \rangle \\
&& \frac{i}{E_p-q_2 - \frac{{\bs p}^2_{\ma{rel}}}{M} - \frac{({\bs p}_{\ma{cm}}-{\bs q}_2)^2}{4M} + i\epsilon}
\\ \nn
i\ml{M}_{(f)} \equiv i\ml{M}^{\mu\nu}_{(f)} (\epsilon_1^*)_\mu (\epsilon_2)_\nu &=& g^2V_A\sqrt{\frac{T_F}{N_c}} f^{abc}(\epsilon_1^*)^0 \Big[ (q_2)^0 (\epsilon_2)^i - (q_2)^i (\epsilon_2)^0 \Big]
\langle \Psi_{{\bs p}_\ma{rel}} | r^i |  \psi_{nl}  \rangle \\
&& \frac{i}{E_p-q_1 - \frac{{\bs p}^2_{\ma{rel}}}{M} - \frac{({\bs p}_{\ma{cm}}-{\bs q}_1)^2}{4M} + i\epsilon}\,.
\ee
We show the Ward identity by replacing $(\epsilon_1)_\mu$ with $(q_1)_\mu$:
\be
i(q_1)_\mu \ml{M}^{\mu\nu}_{(c)}  (\epsilon_2)_\nu &=& i g^2V_A\sqrt{\frac{T_F}{N_c}} f^{abc}  \Big[ - q^0 (\epsilon_2)^i + q^i (\epsilon_2)^0 \Big] \langle \Psi_{{\bs p}_\ma{rel}} | r^i |  \psi_{nl}  \rangle\\
i(q_1)_\mu \ml{M}^{\mu\nu}_{(d)}  (\epsilon_2)_\nu &=& i g^2V_A\sqrt{\frac{T_F}{N_c}} f^{abc}  \Big[  (q_1)^0 (\epsilon_2)^i - (q_1)^i (\epsilon_2)^0 \Big] \langle \Psi_{{\bs p}_\ma{rel}} | r^i |  \psi_{nl}  \rangle\\
i(q_1)_\mu \ml{M}^{\mu\nu}_{(e)}  (\epsilon_2)_\nu &=& 0\\ \nn
i(q_1)_\mu \ml{M}^{\mu\nu}_{(f)}  (\epsilon_2)_\nu &=& i g^2V_A\sqrt{\frac{T_F}{N_c}} f^{abc}  \Big[  (q_2)^0 (\epsilon_2)^i - (q_2)^i (\epsilon_2)^0 \Big] \frac{(q_1)^0\langle \Psi_{{\bs p}_\ma{rel}} | r^i |  \psi_{nl}  \rangle}{E_p-q_1 - \frac{{\bs p}^2_{\ma{rel}}}{M} - \frac{({\bs p}_{\ma{cm}}-{\bs q}_1)^2}{4M}} \\
&=& - i g^2V_A\sqrt{\frac{T_F}{N_c}} f^{abc}  \Big[  (q_2)^0 (\epsilon_2)^i - (q_2)^i (\epsilon_2)^0 \Big] \langle \Psi_{{\bs p}_\ma{rel}} | r^i |  \psi_{nl}  \rangle + \ml{O}(v^2)\,,
\ee
where in the last line we have used $E_p = \frac{{\bs p}_{\ma{rel}}^2}{M} + \frac{{\bs p}_{\ma{cm}}^2}{4M}$ and kept only the relative kinetic energy of the octet by neglecting the c.m.~kinetic energy. This is consistent with our power counting. Since $q^\mu = q_1^\mu - q_2^\mu$, the Ward identity is satisfied up to $v^2$-corrections
\be
i(q_1)_\mu \Big[ \ml{M}^{\mu\nu}_{(c)} + \ml{M}^{\mu\nu}_{(d)}+\ml{M}^{\mu\nu}_{(e)}+\ml{M}^{\mu\nu}_{(f)} \Big] (\epsilon_2)_\nu = 0\,.
\ee
Therefore we can compute these diagrams in any gauge we want. In Coulomb gauge, the zero component of the gauge field is not dynamical. So we only need to compute the diagram in Fig.~\ref{subfig:c},
\be \nn
i\ml{M}_{(c)} &=& -g^2V_A\sqrt{\frac{T_F}{N_c}}  (\epsilon_1^*)_\mu (\epsilon_2)_\nu f^{abc}  \Big[  g^{\nu\rho}(q_2-q)^\mu + g^{\rho\mu}(q+q_1)^\nu + g^{\mu\nu}(-q_1-q_2)^\rho  \Big] \\
&&\Big[ \delta_{\rho j} q_0 \frac{i(\delta_{ji} - \hat{q}_j\hat{q}_i)}{(q_0)^2-{\bs q}^2+i\epsilon }  - \delta_{\rho 0} q_i \frac{i}{{\bs q}^2}  \Big]  \langle \Psi_{{\bs p}_\ma{rel}} | r_i |  \psi_{nl}  \rangle\,.
\ee
We define the square of the amplitude magnitude, summed over all color indexes and polarizations:
\be 
\label{eqn:collinear_g}
\sum |\ml{M}_{(c)} |^2 \equiv \sum_{a,b,c}\sum_{\epsilon_1,\epsilon_2}|\ml{M}_{(c)} |^2 &=& \frac{1}{3}g^4V_A^2C_F | 
\langle \Psi_{{\bs p}_\ma{rel}} | {\bs r} |  \psi_{nl}  \rangle|^2 
\Big[ \frac{1+(\hat{q}_1\cdot\hat{q}_2)^2}{{\bs q}^2} (q_1+q_2)^2 \\ \nn
&& + \frac{1 }{((q_0)^2 - {\bs q}^2 + i\epsilon )^2} P^T_{i_1i_2}({\bs q}_1) P^T_{j_1j_2}({\bs q}_2) P^T_{k_1k_2}({\bs q})\\ \nn
&&  \big( g_{j_1k_1}(q_2-q)_{i_1}  + g_{k_1i_1}(q+q_1)_{j_1} + g_{i_1j_1} (-q_1-q_2)_{k_1}  \big)  \\ \nn
&&  \big( g_{j_2k_2}(q_2-q)_{i_2}  + g_{k_2i_2}(q+q_1)_{j_2} + g_{i_2j_2} (-q_1-q_2)_{k_2}  \big)
 \Big]\,,
\ee
where the transverse polarization tensor is defined as $P^T_{ij}({\bs q}) = \delta_{ij}-\hat{q}_i\hat{q}_j$ and $P^T_{00}=P^T_{0i}=P^T_{i0}=0$. As in the process of inelastic scattering with light quarks, the first term in the square bracket is infrared safe because of the finite binding energy. The second term is collinear divergent when the momenta of the incoming and outgoing gluons are in the same direction. In that case, the transferred gluon is on-shell. As will be shown in Section~\ref{subsect: lmno}, the interference between the diagrams in Figs.~\ref{subfig:a} and~\ref{subfig:m} will cancel this divergence.

\subsubsection{Contributions from diagrams \ref{subfig:g}, \ref{subfig:h}, \ref{subfig:i}, \ref{subfig:j} and \ref{subfig:k}}
\label{subsect: ghijk}
The diagrams in Figs.~\ref{subfig:g}, \ref{subfig:h}, \ref{subfig:i}, \ref{subfig:j} and \ref{subfig:k} describe the processes of $l+\bar{l}+H\leftrightarrow Q+\bar{Q}$ and $g+g+H\leftrightarrow Q+\bar{Q}$. They can be computed similarly as in Sections~\ref {subsect: b} and \ref{subsect: cdef}. However, their contributions to the dissociation and recombination in the Boltzmann equation are much smaller than those from Figs.~\ref{subfig:b}, \ref{subfig:c}, \ref{subfig:d}, \ref{subfig:e} and \ref{subfig:f} because of the limited phase space of the incoming particles. In Coulomb gauge, we only need to consider Figs.~\ref{subfig:g} and \ref{subfig:h}. The energy transferred via the internal gluon is fixed by $q^0=|E_{nl}| + {\bs p}_\ma{rel}^2/M$ and $p_\ma{rel}\sim Mv$ otherwise the dipole transition between wavefunctions $|\langle \Psi_{{\bs p}_\ma{rel}} | r_i |  \psi_{nl}  \rangle|^2$ vanishes. The phase spaces constrained by $p_1+p_2 = q^0$ in Fig.~\ref{subfig:g} and $q_1+q_2 = q^0$ in Fig.~\ref{subfig:h} are much smaller than those of $p_1-p_2 = q^0$ in Fig.~\ref{subfig:b} and $q_1-q_2 = q^0$ in Fig.~\ref{subfig:c}. The suppression of processes with two incoming light quarks or gluons has been noted before \cite{Song:2012at}.

\subsubsection{Contributions from diagrams \ref{subfig:l}, \ref{subfig:m}, \ref{subfig:n} and \ref{subfig:o} }
\label{subsect: lmno}
These diagrams are the one-loop corrections of the gluon propagator. If resummed, they will give a thermal mass to the in-medium gluon. The loop correction is at the order $g^2$ so the whole diagram is at the order $g^3r$. Their interference with the diagram in Fig.~\ref{subfig:a} will give contributions equivalent to amplitudes at the order $g^2r$. We will show the interference cancels the collinear divergence in Eqs.~(\ref{eqn:collinear_q}) and (\ref{eqn:collinear_g}). Thus there is no need to resum these diagrams here.

In Coulomb gauge,
\be
i\ml{M}_{(a)} &=& gV_A\sqrt{\frac{T_F}{N_c}}q_0\epsilon^*_i  \langle \Psi_{{\bs p}_\ma{rel}} | r_i |  \psi_{nl}  \rangle \delta^{ab} \\
\label{eqn:M_l}
i\ml{M}_{(l)} &=&  gV_A\sqrt{\frac{T_F}{N_c}}\epsilon^*_i  \langle \Psi_{{\bs p}_\ma{rel}} | r_k |  \psi_{nl}  \rangle \delta^{ab} \Big[ i\Pi^{(l)}_{ij} \frac{iq_0(\delta_{jk}-\hat{q}_j\hat{q}_k)}{q_0^2-{\bs q}^2 + i\epsilon}   - i\Pi^{(l)}_{i0}\frac{iq_k}{{\bs q}^2}    \Big]\,,
\ee
where we set $q_0=|\bs q| \equiv q$ for the on-shell massless particle and $\Pi^{(l)}_{\mu\nu}$ is the time-ordered gluon polarization tensor contributed from the fermion loop. The time-ordered gluon propagator and polarization at finite temperature cannot be directly obtained by analytically continuing the imaginary time propagator and polarization. The time-ordered propagators and polarizations can only be obtained from the retarded and advanced ones via the relation
\be
D^T(q_0,{\bs q})   &=& \frac{1}{2}\big( D^R(q_0,{\bs q}) + D^A(q_0,{\bs q})    \big)  + \big( \frac{1}{2}+n_B(q_0) \big) \big( D^R(q_0,{\bs q}) - D^A(q_0,{\bs q}) \big)\\
\Pi^T(q_0,{\bs q}) &=& \frac{1}{2}\big( \Pi^R(q_0,{\bs q}) + \Pi^A(q_0,{\bs q}) \big)  + \big( \frac{1}{2}+n_B(q_0) \big) \big( \Pi^R(q_0,{\bs q}) - \Pi^A(q_0,{\bs q}) \big)\,.
\ee
We will focus on the first term $\frac{D^R(q_0,{\bs q}) + D^A(q_0,{\bs q})}{2}$ because this term contributes to the dissociation rate when the gluon is on-shell. The second term contributes to the dissociation rate when the gluon has space-like momentum, which corresponds to the inelastic scattering and has been accounted for above. For a more detailed discussion on this, see Ref.~\cite{Brambilla:2013dpa}. The fermion loop gives,
\be
i\Pi^{(l)}_{\mu\nu}(q_0, {\bs q}) = -g^2T_F\sum_{\ma{flavor}}\int\frac{\diff^4 l}{(2\pi)^4} \frac{\Tr{(\gamma_\mu(\slashed{l}+\slashed{q})\gamma_\nu \slashed{l}})}{ (l_0^2-{\bs l}^2)^2 ((l_0+q_0)^2-({\bs l}+{\bs q})^2)^2}\,.
\ee
In the imaginary formalism of thermal field theory, the integral over $l_0$ is a summation in Matsubara frequency. After the summation
\be \nn
\label{eqn:pi_q}
\Pi_{ij}^{(l)}(q_0=q, {\bs q}) &=& g^2T_F\sum_{\ma{flavor}} \int\frac{\diff^3 l}{(2\pi)^3} \frac{\Tr{(\gamma_i (\slashed{l}+\slashed{q})\gamma_j \slashed{l}})}{4E_1E_2} \\ \nn
&& \Big[  (1-n_F(E_1)-n_F(E_2) ) \Big(\frac{1}{q-E_1-E_2} - \frac{1}{q+E_1+E_2} \Big) \\
&& -  (n_F(E_1)-n_F(E_2) ) \Big(\frac{1}{q+E_1-E_2} - \frac{1}{q-E_1+E_2} \Big)
\Big]\,,
\ee
where $E_1 = |{\bs l}+{\bs q}|$, $E_2 = |{\bs l}|$. Here we only need $\frac{\Pi^R(q_0,{\bs q}) + \Pi^A(q_0,{\bs q})}{2} = \Re{\Pi^R(q_0,{\bs q})}$ and we do not need to analytically continue. We can just plug it into (\ref{eqn:M_l}).
To see the cancellation of collinear divergence, we define the interference term summed over colors and gluon polarizations
\be
\sum (\ml{M}_{(a)}^*\ml{M}_{(l)} + \ml{M}_{(a)} \ml{M}_{(l)}^*) &\equiv& \sum_\epsilon \sum_{a,b} (\ml{M}_{(a)}^*\ml{M}_{(l)} + \ml{M}_{(a)} \ml{M}_{(l)}^*) \\
&=& -\frac{2}{3}g^2V_A^2C_F q_0^2  | \langle \Psi_{{\bs p}_\ma{rel}} | {\bs r} |  \psi_{nl}  \rangle |^2  \frac{\delta_{ij}-\hat{q}_i\hat{q}_j}{q_0^2-{\bs q}^2 } \Pi^{(l)}_{ij}\,,
\ee
and consider the following integral that is used in the dissociation in the case of two flavors (up and down quarks)
\be
I_1 \equiv  \int\frac{\diff^3 q}{2q (2\pi)^3} n_B(q)\sum (\ml{M}_{(a)}^*\ml{M}_{(l)} + \ml{M}_{(a)} \ml{M}_{(l)}^*)\,.
\ee
We focus on the term with $(n_F(E_1)-n_F(E_2) )$ in the square bracket in Eq. (\ref{eqn:pi_q}). Under a change of variables ${\bs p}_1 = {\bs l}+{\bs q}$, ${\bs p}_2 = {\bs l}$
\be
I_1 &\equiv&  \int\frac{\diff^3 p_1}{2p_1 (2\pi)^3} \int\frac{\diff^3 p_2}{2p_2 (2\pi)^3} n_B(|{\bs p}_1-{\bs p}_2|) (n_F(p_1) - n_F(p_2)) \\ \nn
&& \frac{16}{3}g^4V_A^2T_FC_F\frac{2(p_1-p_2)|{\bs p}_1-{\bs p}_2|}{((p_1-p_2)^2-({\bs p}_1-{\bs p}_2)^2)^2} (p_1p_2-{\bs p}_1\cdot\hat{q}{\bs p}_2\cdot\hat{q}) + \cdots\,,
\ee
where $\cdots$ means the first term in the square bracket in Eq.~(\ref{eqn:pi_q}). In the collinear limit, ${\bs p}_1$ and ${\bs p}_2$ are in the same direction, so $p_1-p_2 = |{\bs p}_1-{\bs p}_2|$. With $n_B(p_1-p_2) (n_F(p_1) - n_F(p_2)) = -n_F(p_1)(1-n_F(p_2))$, one immediately sees $I_1$ cancels the collinear divergence in
\be
\int \frac{\diff^3 p_1}{2p_1(2\pi)^3} \frac{\diff^3 p_2}{2p_2(2\pi)^3}  n_F(p_1)(1-n_F(p_2)) \sum |\ml{M}_{(b)}|^2\,,
\ee
where $|\ml{M}_{(b)}|^2$ is given in Eq.~(\ref{eqn:collinear_q}). This is for the dissociation contribution from $|\ml{M}_{(b)}|^2$. The cancellation of the collinear divergence in the recombination can be shown similarly.

We next consider the interference between the diagrams in Figs.~\ref{subfig:a} and \ref{subfig:m}. The amplitude $i \ml{M}_{(m)}$ is exactly the same as $i\ml{M}_{(l)}$ under the replacement of $\Pi^{(l)}$ with $\Pi^{(m)}$. $\Pi^{(m)}_{\mu\nu}$ is the gluon polarization tensor in Fig.~\ref{subfig:m}. After summing over the Matsubara frequencies
\be \nn
\Pi^{(m)}_{ij}(q_0=q,{\bs q}) &=& \frac{1}{2}g^2T_A \int\frac{\diff^3l }{(2\pi)^3} \frac{1}{4E_1E_2} P^T_{i_1i_2}({\bs l}+{\bs q}) P^T_{j_1j_2}({\bs l})\\ \nn
&& \Big[ (1+n_B(E_1)+n_B(E_2)) \Big(  \frac{1}{q+E_1+E_2} - \frac{1}{q-E_1-E_2} \Big) \\ \nn
&& -(n_B(E_1)-n_B(E_2)) \Big( \frac{1}{q+E_1-E_2} - \frac{1}{q-E_1+E_2}   \Big)   \Big]  \\ \nn
&& \Big[ g_{j_1i}(l-q)_{i_1}  + g_{ii_1}(2q+l)_{j_1} + g_{i_1j_1} (-q-2l)_{i}   \Big] \\
&& \Big[ g_{j_2j}(l-q)_{i_2}  + g_{ji_2}(2q+l)_{j_2} + g_{i_2j_2} (-q-2l)_{j}  \Big]\,,
\ee
where $T_A=N_c$, $E_1 = |{\bs l}+{\bs q}|$ and $E_2 = |{\bs l}|$. We will focus on the term with $(n_B(E_1)-n_B(E_2))$ and neglect the other term. Under a change of variables ${\bs q}_1={\bs l}+{\bs q}$, ${\bs q}_2={\bs l}$, we can show that the collinear divergence of the integral
\be
I_2 \equiv  \int\frac{\diff^3 q}{2q (2\pi)^3} n_B(q)\sum (\ml{M}_{(a)}^*\ml{M}_{(m)} + \ml{M}_{(a)} \ml{M}_{(m)}^*)\,,
\ee
cancels the collinear divergence in 
\be
\int \frac{\diff^3 q_1}{2q_1(2\pi)^3} \frac{\diff^3 q_2}{2q_2(2\pi)^3}  n_B(q_1)(1+n_B(q_2)) \sum |\ml{M}_{(c)}|^2\,,
\ee
by the virtual of $n_B(q_1-q_2)(n_B(q_1)-n_B(q_2)) = -n_B(q_1)(1+n_B(q_2))$. Cancellation of the divergence in the recombination process can be similarly shown.

\subsubsection{Contributions from diagrams \ref{subfig:p}, \ref{subfig:q}, \ref{subfig:r} and \ref{subfig:s}}
\label{subsect: pqrs}
These diagrams are the one-loop corrections to the dipole interaction between the singlet and the octet. The correction is at the order $g^3r$ but its interference with the diagram in Fig.~\ref{subfig:a} gives contributions equivalent to amplitudes at the order $g^2r$. In Lorentz gauge and dimensional regularization $d=4-\epsilon$ (the $\epsilon$ in the dimensional regularization should be distinguished from the incoming gluon polarization $\epsilon^{*\mu}$)
\be
i\ml{M}_{(p)} &=& - \frac{3g^3}{16\pi^2\epsilon} V_A T_A\sqrt{\frac{T_F}{N_c}} \delta^{ad} \langle \Psi_{{\bs p}_\ma{rel}} | r^i |  \psi_{nl}  \rangle 
(\epsilon^{*0} q^i - \epsilon^{*i} q^0 ) +\cdots\\
i\ml{M}_{(q)} &=&  \frac{3g^3}{16\pi^2\epsilon}  V_A T_A\sqrt{\frac{T_F}{N_c}} \delta^{ad} \langle \Psi_{{\bs p}_\ma{rel}} | r^i |  \psi_{nl}  \rangle 
(\epsilon^{*0} q^i - \epsilon^{*i} q^0 ) +\cdots\\
i\ml{M}_{(r)} &=& 0 +\cdots\\
i\ml{M}_{(s)} &=& 0 +\cdots\,,
\ee
where $T_A=N_c$ and the off-shell scheme has been used to extract the logarithmic divergence. Only the terms with the $\epsilon$ poles are shown. Finite terms and corrections at higher orders in $v^2$ are omitted. Therefore
\be
i\ml{M}_{(p)} + i\ml{M}_{(q)} + i\ml{M}_{(r)} + i\ml{M}_{(s)} = \frac{0}{\epsilon} + \cdots\,,
\ee
which means the dipole interaction term between the singlet and octet is independent of scale at the one-loop level
\be
\frac{\diff}{ \diff \mu} V_A(\mu) = 0\,.
\ee
This has been already noted in Ref. \cite{Pineda:2000gza}. From the matching condition, Eq.~(\ref{eqn:match}), we may set $V_A=1$ in the following, no matter the scale involved.

\subsubsection{Contributions from diagram \ref{subfig:t} }
\label{subsect: t}
The diagram in Fig.~\ref{subfig:t} describes the one-loop correction to the octet propagator at the order $g^2$. The whole diagram is at the order $g^3r$ but its interference with the diagram \ref{subfig:a} is equivalent to an amplitude at the order $g^2r$. The one-loop correction to the octet propagator is given by
\be
L_{o} \equiv g^2f^{abc}f^{cbd} \int\frac{\diff^4l}{(2\pi)^4} \frac{-i}{l_0^2-{\bs l}^2 + i\epsilon} \frac{i}{E_p - l_0 - \frac{{\bs p}_\ma{rel}^2}{M} - \frac{({\bs p}_\ma{cm}-{\bs l})^2}{4M} + i\epsilon}\,,
\ee
where $E_p = \frac{{\bs p}_\ma{rel}^2}{M} + \frac{ {\bs p}_\ma{cm}^2 }{4M} $ is the energy of the external octet field.
We first integrate over $l_0$ by closing the contour in the lower half plane
\be
L_{o} &=& i g^2N_c \delta^{ad} \int\frac{\diff^3l}{(2\pi)^3} \frac{1}{2l} \frac{1}{l+   (l^2 - 2 {\bs l} \cdot {\bs P}_\ma{cm})/(4M) } \\
&=& \frac{ig^2N_c \delta^{ad} M}{4\pi^2 P_\ma{cm}} \int \diff l \ln{\frac{l+4M+2P_\ma{cm}}{l+4M-2P_\ma{cm}}}\,.
\ee
If we expand the integrand in powers of $1/M$ (in our power counting, $v^2 \sim \frac{P_\ma{cm}}{M}$) and use dimensional regularization\footnote{Similar argument has been used in Ref.~\cite{Bodwin:1994jh}. See the appendix therein.}, we find
\be
L_{o} = 0 + \ml{O}(v^2)\,.
\ee
The power divergence is proportional to $v^2$ and neglected here, consistent with the power counting.

\subsubsection{Summary}
To write the dissociation and recombination terms in the Boltzmann equations explicitly, we define $\ml{F}^\pm_{nls(b)}$ and $\ml{F}^\pm_{nls(c)}$ as
\be \nn
\ml{F}^+_{nls(b)} &\equiv&  g_+ \int \frac{\diff^3 k}{(2\pi)^3}  \frac{\diff^3 p_{Q}}{(2\pi)^3}  \frac{\diff^3 p_{\bar{Q}} }{(2\pi)^3} \frac{\diff^3 p_1}{2p_1(2\pi)^3} \frac{\diff^3 p_2}{2p_2(2\pi)^3}  n_F(p_2)(1-n_F(p_1)) f_Q({\bs x}_Q, {\bs p}_Q,t) f_{\bar{Q}} ({\bs x}_{\bar{Q}}, {\bs p}_{\bar{Q}},t) \\
&& (2\pi)^4\delta^3({\bs k} + {\bs p}_1 - {\bs p}_{\ma{cm}} - {\bs p}_2) \delta(-|E_{nl}|+p_1-\frac{p^2_{\ma{rel}}}{M}-p_2)  \sum | \ml{M}_{(b)} |^2   \\ \nn
\ml{F}^-_{nls(b)}  &\equiv&  \int \frac{\diff^3 k}{(2\pi)^3}  \frac{\diff^3 p_{\ma{cm}}}{(2\pi)^3}  \frac{\diff^3 p_{\ma{rel}} }{(2\pi)^3} \frac{\diff^3 p_1}{2p_1(2\pi)^3} \frac{\diff^3 p_2}{2p_2(2\pi)^3}  n_F(p_1)(1-n_F(p_2))  f_{nls}({\bs x}, {\bs k}, t) \\
&& (2\pi)^4\delta^3({\bs k} + {\bs p}_1 - {\bs p}_{\ma{cm}} - {\bs p}_2) \delta(-|E_{nl}|+p_1-\frac{p^2_{\ma{rel}}}{M}-p_2)  \sum | \ml{M}_{(b)} |^2 \\ \nn
\ml{F}^+_{nls(c)} &\equiv&  g_+ \int \frac{\diff^3 k}{(2\pi)^3}  \frac{\diff^3 p_{Q}}{(2\pi)^3}  \frac{\diff^3 p_{\bar{Q}} }{(2\pi)^3} \frac{\diff^3 q_1}{2q_1(2\pi)^3} \frac{\diff^3 q_2}{2q_2(2\pi)^3}  n_B(q_2)(1+n_B(q_1)) f_Q({\bs x}_Q, {\bs p}_Q,t) f_{\bar{Q}} ({\bs x}_{\bar{Q}}, {\bs p}_{\bar{Q}},t) \\
&& (2\pi)^4\delta^3({\bs k} + {\bs q}_1 - {\bs p}_{\ma{cm}} - {\bs q}_2) \delta(-|E_{nl}|+q_1-\frac{p^2_{\ma{rel}}}{M}-q_2)  \sum | \ml{M}_{(c)} |^2   \\ \nn
\ml{F}^-_{nls(c)}  &\equiv&  \int \frac{\diff^3 k}{(2\pi)^3}  \frac{\diff^3 p_{\ma{cm}}}{(2\pi)^3}  \frac{\diff^3 p_{\ma{rel}} }{(2\pi)^3} \frac{\diff^3 q_1}{2q_1(2\pi)^3} \frac{\diff^3 q_2}{2q_2(2\pi)^3}  n_B(q_1)(1+n_B(q_2))  f_{nls}({\bs x}, {\bs k}, t) \\
&& (2\pi)^4\delta^3({\bs k} + {\bs q}_1 - {\bs p}_{\ma{cm}} - {\bs q}_2) \delta(-|E_{nl}|+q_1-\frac{p^2_{\ma{rel}}}{M}-q_2)  \sum | \ml{M}_{(c)} |^2 \,,
\ee
The $g_+$ factor and the relation between ${\bs p}_{\ma{cm}}$, ${\bs p}_{\ma{rel}}$ and ${\bs p}_{Q}$, ${\bs p}_{\bar{Q}}$ are defined in Section~\ref{subsect: a}. The collinear divergent parts in the square of amplitudes have been shown to be cancelled by the interference between the tree-level process of gluon absorption/emission and its one-loop corrections. After regularization, we can drop the terms that are originally collinear divergent if they are small, so we can write (by setting $V_A=1$)
\be
\sum | \ml{M}_{(b)} |^2  &=& \frac{16}{3}g^4T_FC_F  |\langle \Psi_{{\bs p}_\ma{rel}} | {\bs r} |  \psi_{nl}  \rangle|^2   \frac{p_1p_2 + {\bs p}_1\cdot {\bs p}_2}{{\bs q}^2} \\
\sum | \ml{M}_{(c)} |^2  &=& \frac{1}{3}g^4C_F | \langle \Psi_{{\bs p}_\ma{rel}} | {\bs r} |  \psi_{nl}  \rangle|^2 
 \frac{1+(\hat{q}_1\cdot\hat{q}_2)^2}{{\bs q}^2} (q_1+q_2)^2\,.
\ee
The dissociation and recombination terms in the Boltzmann equation from the inelastic scattering with light quarks and gluons are given by
\be
\ml{C}_{nls,\ma{inel}}^{\pm}({\bs x}, {\bs p}, t) &=&  \frac{\delta \ml{F}^\pm_{nls(b)}}{\delta{{\bs k}}}  \Big|_{{\bs k}={\bs p}} + \frac{\delta \ml{F}^\pm_{nls(c)}}{\delta{{\bs k}}}  \Big|_{{\bs k}={\bs p}}\,.
\ee
For $\ml{C}_{nls,\ma{inel}}^{+}$, we further require ${\bs x}=\frac{{\bs x}_Q+{\bs x}_{\bar{Q}}}{2}$ as in the case of real gluon absorption. The dissociation rate of the quarkonium state $nls$ from the inelastic scattering is given by
\be
\Gamma_{nls,\ma{inel}}^{\ma{disso}}({\bs x}, {\bs p}, t) \equiv \frac{\ml{C}_{nls,\ma{inel}}^{-}({\bs x}, {\bs p}, t)}{ f_{nls}({\bs x}, {\bs p}, t) }\,.
\ee
The recombination rate of a heavy quark into the quarkonium state $nls$ surrounded by heavy antiquarks with the distribution $f_{\bar{Q}} ({\bs x}_{\bar{Q}}, {\bs p}_{\bar{Q}},t)$ is given by
\be
\Gamma_{nls,\ma{inel}}^{\ma{recom}}({\bs x}, {\bs p}, t) \equiv \frac{1}{f_Q({\bs x}_Q, {\bs p}_Q,t) }  \frac{\delta (\ml{F}^+_{nls(b)}+\ml{F}^+_{nls(c)}) }{\delta{{\bs p}_Q}} \Big|_{{\bs x}_Q={\bs x},\ {\bs p}_Q={\bs p}}\,.
\ee

\section{Diffusion and energy loss}
\label{sect:diffuse}
\begin{figure}
    \centering
    \begin{subfigure}[b]{0.33\textwidth}
        \centering
        \includegraphics[height=1.25in]{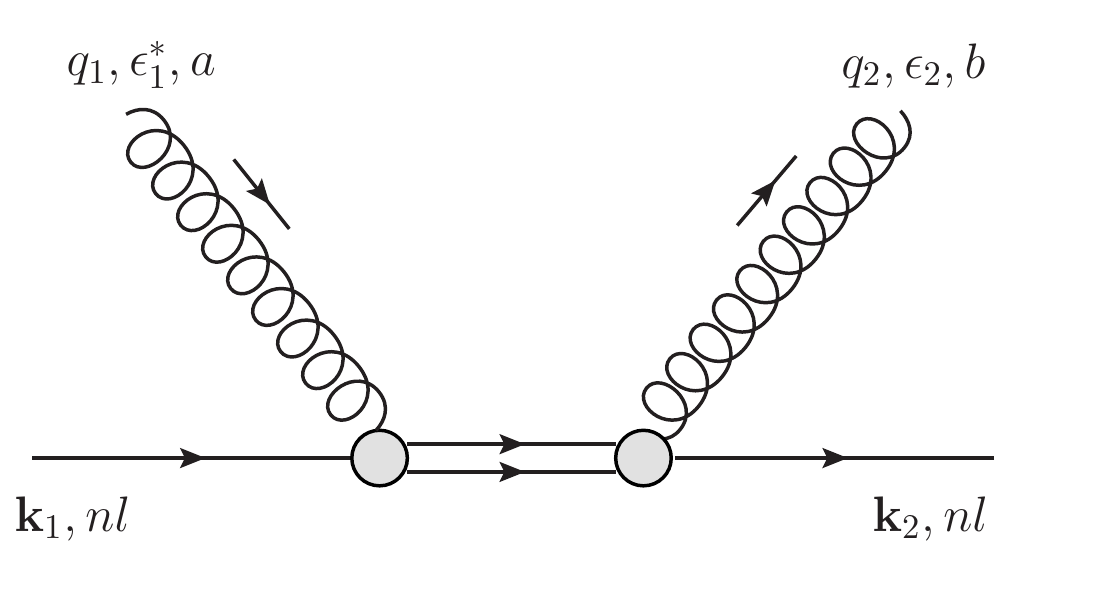}
        \caption{}\label{subfig:elastic1}
    \end{subfigure}%
    ~ 
    \begin{subfigure}[b]{0.33\textwidth}
        \centering
        \includegraphics[height=1.25in]{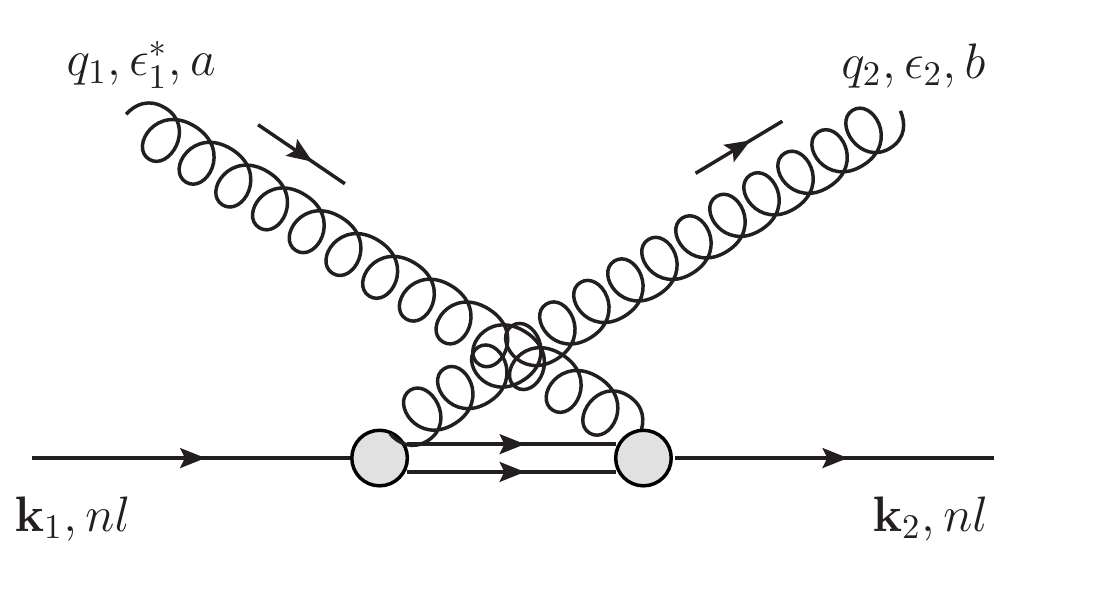}
        \caption{}\label{subfig:elastic2}
    \end{subfigure}%
    ~
    \begin{subfigure}[b]{0.33\textwidth}
        \centering
        \includegraphics[height=1.25in]{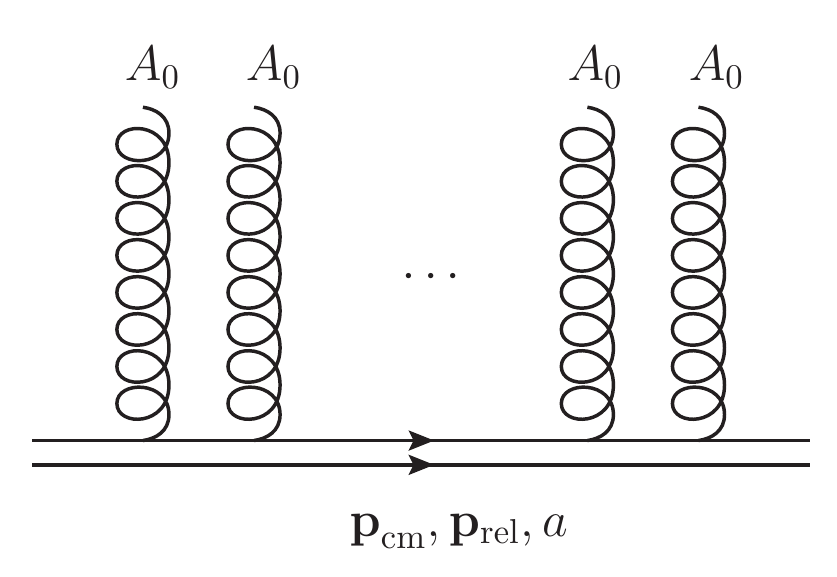}
        \caption{}\label{subfig:diffuse}
    \end{subfigure}%
    \caption{Feynman diagrams contributing to the quarkonium diffusion term in the Boltzmann equation. First two diagrams are the processes at the order $g^2r^2$. The last diagram is schematic and shows the virtual octet propagator can in principle obtain an infinite series of momentum ``kicks" from the medium.}
    \label{fig:diagrams2}
\end{figure}

The quarkonium diffusion cannot happen at the order $r$, because the singlet has to turn to an octet at this order. The diffusion process starts to happen at the order $r^2$ because the singlet can turn to an octet and then become a singlet again. At the order $g^2r^2$, contributing diagrams are shown in Fig.~\ref{subfig:elastic1} and \ref{subfig:elastic2}. They are from the dipole vertex at the second order in perturbation theory.\footnote{At the order $g^2r^2$, we need to consider new terms that show up in the Lagrangian in the multipole expansion. A quadrupole term of the form $g^2S^\dagger r_ir_j S$ contracted with $E_{i}$ or $A_{i}$ can contribute at the first order in perturbation theory. However, such a term has a vanishing matching coefficient \cite{Brambilla:2003nt}.} The amplitudes satisfy the Ward identity by virtue of Eq.~(\ref{eqn:ward1}). In Coulomb gauge, the amplitudes of diagrams \ref{subfig:elastic1} and \ref{subfig:elastic2} are
\be \nn
i\ml{M} &=& -g^2\frac{T_F}{N_c}\delta^{ab}(\epsilon^*_1)_i(\epsilon_2)_j q_1q_2  \\
&& \int\frac{\diff^3 p_\ma{rel}}{(2\pi)^3} \bigg(
\frac{\langle \psi_{nl} | r_j | \Psi_{{\bs p}_\ma{rel}} \rangle   \langle  \Psi_{{\bs p}_\ma{rel}} |r_i | \psi_{nl} \rangle}{q_1-|E_{nl}| - \frac{{\bs p}_\ma{rel}^2}{M} + i\epsilon }  
+ \frac{\langle \psi_{nl} | r_j | \Psi_{{\bs p}_\ma{rel}} \rangle \langle  \Psi_{{\bs p}_\ma{rel}} |r_i | \psi_{nl} \rangle }{-q_1-|E_{nl}| - \frac{{\bs p}_\ma{rel}^2}{M} + i\epsilon }    \bigg)
\,.
\ee
When $q_1 \ge |E_{nl}|$, the first term in the big bracket has a pole. At the pole, the term becomes imaginary. Physically, this happens when the intermediate octet state becomes on-shell so the process becomes the quarkonium dissociation. Therefore we should take the principal value of the integral $\ml{P}\int\diff^3 p_{\ma{rel}}$. One can show the principal value is well-defined, i.e., the divergent contributions from both sides of the pole cancel out. 

During the lifetime of the virtual octet, its momentum may change due to a number of collisions that transfer a small momentum, as depicted in Fig.~\ref{subfig:diffuse}. These processes are at the order $r^0$, so not suppressed by the multipole expansion. The virtual octet diffuses as if it were an open heavy quark. Since the contributions cancel out near the pole of the octet propagator, the octet behaves like a state with lifetime $\Delta\tau\sim\frac{1}{Mv^2}$. The rate of transferring the square of momentum is about $\alpha_s^2T^3$ \cite{Moore:2004tg}. So the square of momentum transferred during its lifetime is about $\frac{\alpha_s^2T^3}{Mv^2}\lesssim\alpha_s^2T^2$ since we assume $T\lesssim Mv^2$. The momentum transferred is about $\alpha_s T$. The c.m.~momentum of the octet is at least $q_1\sim T \gg \alpha_sT$. So the effect from the virtual octet diffusion is small and there is no need to resum $gA_0$ into the virtual octet.

We define the square of the total amplitude, summed over colors and polarizations of gluons
\be \nn
\sum |\ml{M}|^2 &\equiv& \sum_{a,b}\sum_{\epsilon_1,\epsilon_2} |\ml{M}|^2 \\
&=& \frac{4}{9}g^4\frac{T_F}{N_c}C_F |\vec{\epsilon}_1^{\ *}\cdot \vec{\epsilon}_2|^2
q_1^2q_2^2 \bigg( \ml{P}\int\frac{\diff^3 p_\ma{rel}}{(2\pi)^3} \frac{|\langle  \Psi_{{\bs p}_\ma{rel}}  | {\bs r} | \psi_{nl}  \rangle|^2 (|E_{nl}|+\frac{{\bs p}^2_\ma{rel}}{M})}{(|E_{nl}|+\frac{{\bs p}^2_\ma{rel}}{M})^2-q_1^2} \bigg)^2\,.
\ee

\begin{figure}
    \centering
    \begin{subfigure}[b]{0.5\textwidth}
        \centering
        \includegraphics[height=2.in]{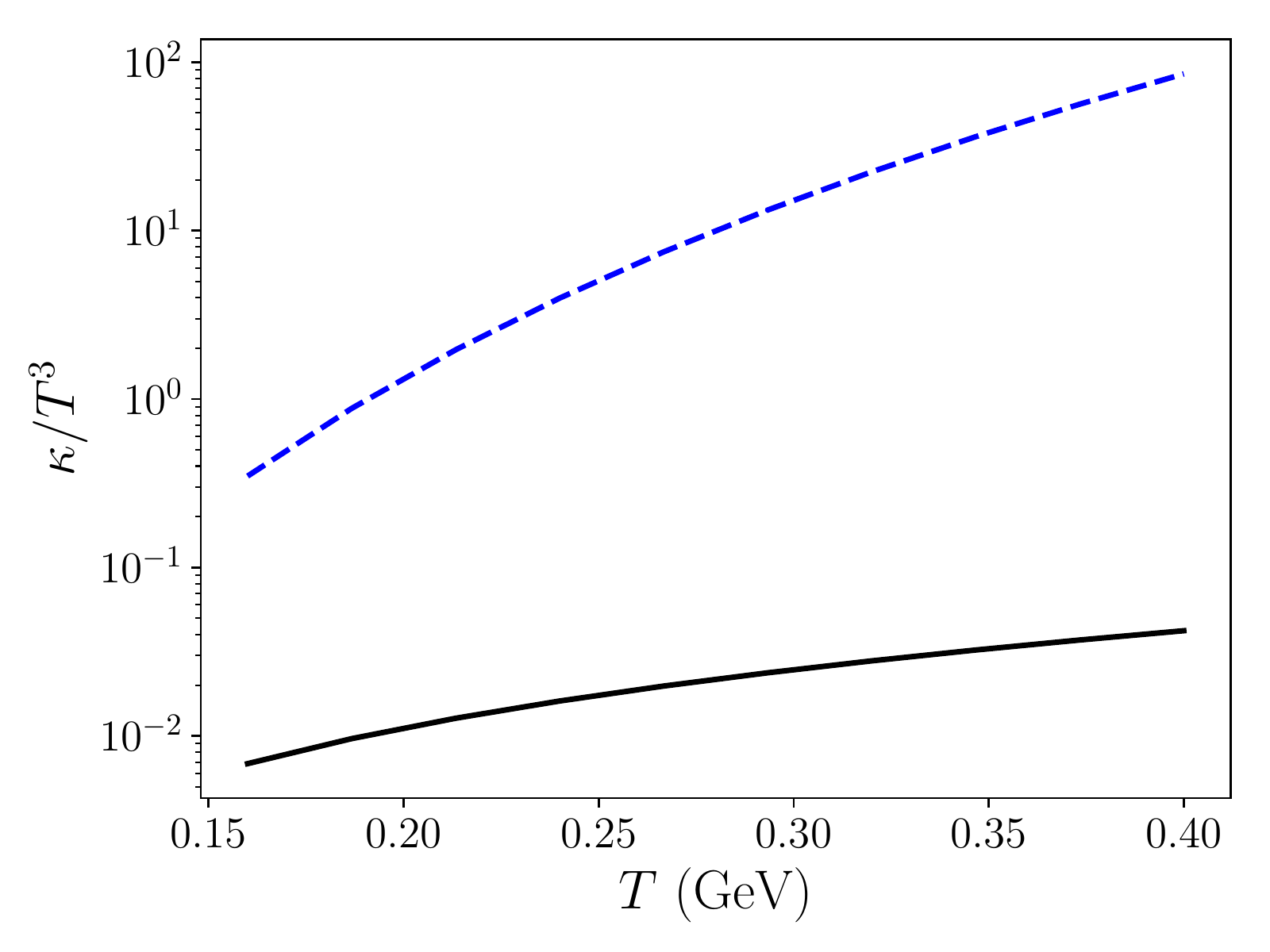}
        \caption{}\label{subfig:kappa}
    \end{subfigure}%
    ~ 
    \begin{subfigure}[b]{0.5\textwidth}
        \centering
        \includegraphics[height=2.in]{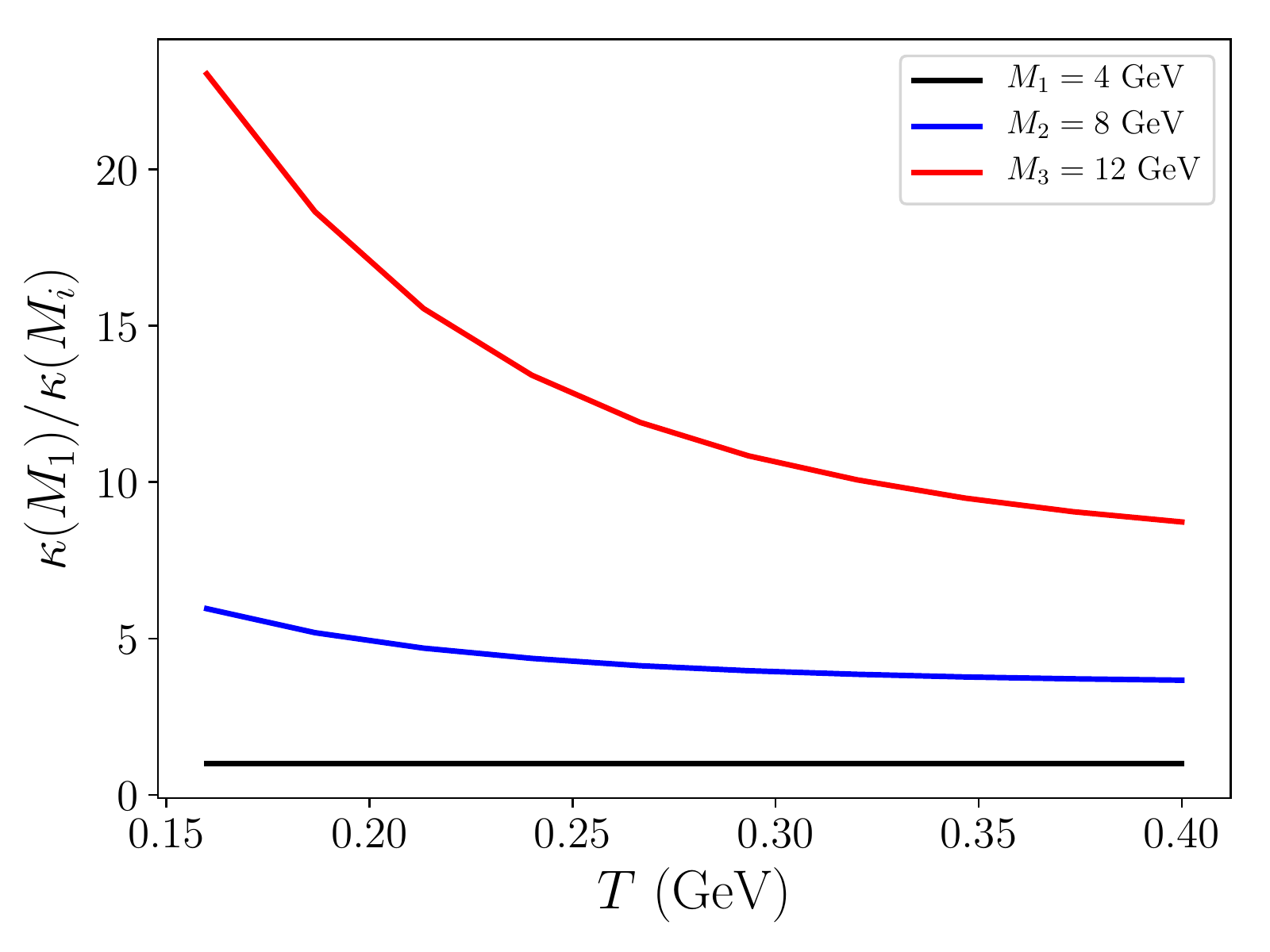}
        \caption{}\label{subfig:kappa_M}
    \end{subfigure}%
    ~
    \caption{$\Upsilon$(1S) diffusion coefficient $\kappa$: (a) as a function of temperature, the solid line is the exact result from Eq.~(\ref{eqn:kappa}) while the dashed line is the approximate result from Eq.~(\ref{eqn:kappa_appro}); (b) mass dependence, the lower, middle and upper lines correspond to $\kappa(M_1)/\kappa(M_i)$ with $i=1,2,3$ respectively. 
    }
    \label{fig:kappa}
\end{figure}

To write the diffusion term in the Boltzmann equation explicitly, we define
\be \nn
\ml{F}_{nls} &\equiv& \int\frac{\diff^3k_1}{(2\pi)^3} \int\frac{\diff^3k_2}{(2\pi)^3} \int\frac{\diff^3q_1}{2q_1(2\pi)^3} \int\frac{\diff^3q_2}{2q_2(2\pi)^3} n_B(q_1)(1+n_B(q_2)) f_{nls}({\bs x}, {\bs k}_1, t)\\
&&(2\pi)^4\delta^3({\bs k}_1+{\bs q}_1 - {\bs k}_2 - {\bs q}_2) \delta(q_1-q_2)  \sum |\ml{M}|^2 \,.
\ee
The diffusion term in the Boltzmann equation (\ref{eqn:boltzmann}) can be written as
\be
\ml{C}_{nls}({\bs x}, {\bs p}, t) = -\frac{\delta \ml{F}_{nls}}{\delta {\bs k}_1} \Big|_{{\bs k}_1 = {\bs p}} + \frac{\delta \ml{F}_{nls}}{\delta {\bs k}_2} \Big|_{{\bs k}_2 = {\bs p}}\,.
\ee
We can also define the diffusion coefficient as the square of momentum transferred per unit time
\be \nn
3\kappa &\equiv& \int\frac{\diff^3k_2}{(2\pi)^3} \int\frac{\diff^3q_1}{2q_1(2\pi)^3} \int\frac{\diff^3q_2}{2q_2(2\pi)^3} n_B(q_1)(1+n_B(q_2)) \\
&&({\bs q}_1-{\bs q}_2)^2(2\pi)^4\delta^3({\bs k}_1+{\bs q}_1 - {\bs k}_2 - {\bs q}_2) \delta(q_1-q_2)  \sum |\ml{M}|^2\,.
\ee
After some simplications
\be
\label{eqn:kappa}
\kappa = \frac{32}{729\pi^5}\alpha_s^2\int \diff q \,q^8 n_B(q)(1+n_B(q))  \bigg( \ml{P}\int  \diff p_\ma{rel} \frac{p_\ma{rel}^2 
|\langle \Psi_{{\bs p}_\ma{rel}} | {\bs r} | \psi_{nl} \rangle|^2
 (|E_{nl}|+\frac{{\bs p}^2_\ma{rel}}{M})}{(|E_{nl}|+\frac{{\bs p}^2_\ma{rel}}{M})^2-q^2} \bigg)^2\,.
\ee
For 1S state, if the $q^2$ in the denominator inside the integral over $p_\ma{rel}$ is neglected and the octet relative wavefunction is a plane wave, one can show
\be
\label{eqn:kappa_appro}
\kappa'(\ma{1S}) = \frac{T^3(\pi T a_B)^6}{N_c^2} \frac{50176\pi}{1215} \frac{2}{C_F^2}\,,
\ee
where the Bohr radius $a_B = \frac{2}{\alpha_sC_F M}$. If one takes the large-$N_c$ approximation, $C_F=3/2$, and multiplies the expression (\ref{eqn:kappa_appro}) by a factor of $9/8$ (because when we sum over colors, there is a factor of $8$, in large-$N_c$, the factor is $9$), then Eq.~(\ref{eqn:kappa_appro}) agrees with the previous estimate using perturbative calculations in another effective field theory where the octet is integrated out \cite{Dusling:2008tg}. The approximate result scales as $\kappa' \propto T^9$. Both the exact result, Eq.~(\ref{eqn:kappa}), and the approximate result, Eq.~(\ref{eqn:kappa_appro}), are shown in Fig.~\ref{fig:kappa} for $\Upsilon$(1S) with $M=4.65$ GeV and $\alpha_s = 0.3$. The two results differ by two to three orders of magnitude. The approximate result Eq.~(\ref{eqn:kappa_appro}) is only valid when $T\ll Mv^2$ so one can neglect the $q^2\sim T^2$ in the denominator.\footnote{In fact, if one expands the integrand of (\ref{eqn:kappa}) in powers of $\frac{q^2}{(|E_{nl}|+{\bs p}^2_\ma{rel}/M)^2}$, one obtains an asymptotic series.} However, for real QGP, $T\gtrsim 160$ MeV, it is not a good approximation even for the bottom quark with $Mv^2\sim 450$ MeV. One should also notice that the typical value of $q$ can be a few times larger than $T$ because of the high power $q^8$ in the phase space integral. This makes the approximation less valid. 
At high temperatures $Mv \gg T \gg Mv^2$ (the first inequality assures our power counting), the $q^2$ in the denominator dominates over $Mv^2$ and we expect $\kappa\propto T^5$. Furthermore, if we assume the octet relative wavefunction is a plane wave,
\be
|\langle \Psi_{{\bs p}_\ma{rel}} | {\bs r} | \psi_{1S} \rangle|^2 = 1024\pi \frac{a_B^5(a_Bp_\ma{rel})^2}{(1+(a_Bp_\ma{rel})^2)^6}\,,
\ee
we expect $\kappa(\ma{1S}) \propto (Ma_B^2)^2  \propto M^{-2}$ at high temperatures. 
The mass dependence of $\kappa$ is also plotted for three difference heavy quark masses. At high temperature the mass scaling is approximately valid.

In principle, a quarkonium has two ways to lose energy inside QGP. One way is the elastic scattering or diffusion. The other way is to dissociate first, lose energy as an unbound heavy quark antiquark pair and then recombine later. The former mechanism only works when the quarkonium is a well-defined bound state inside QGP. So it only makes sense when the temperature is below the quarkonium melting temperature. As shown in Fig.~\ref{fig:kappa}, the rate of momentum transfer due to the diffusion is very slow for $\Upsilon$(1S), compared with that of open heavy quarks ($\kappa/T^3$ of heavy quarks is on the order of $1$ or $10$ \cite{Cao:2018ews}). This is also true for $J/\psi$, because we expect its diffusion coefficient is $10$ times larger than that of $\Upsilon$(1S) from the mass scaling. But $J/\psi$ has lower melting temperature, so the diffusion coefficient would probably only make sense when $T$ is below $300$ MeV. Therefore the latter mechanism (dissociation followed by energy loss and recombination) probably dominates the quarkonium energy loss, even though not every quarkonium finally observed has to go through this sequence of processes. Some of the primordially produced quarkonia may survive the in-medium evolution and lose almost no energy.

\section{Conclusion}
\label{sect:conclusion}
In this paper we calculated the dissociation, recombination and diffusion terms in the Boltzmann transport equation of quarkonium. We considered the processes of gluon absorption/emission, inelastic scattering and elastic scattering with medium constituents. We computed scattering amplitudes directly in pNRQCD and showed they satisfy the Ward identities. Loop corrections were also considered. The dipole interaction is not running at the one-loop level. The interference between the gluon absorption/emission and its thermal loop corrections cancels the collinear divergence in the inelastic scattering. The inelastic scattering amplitude is infrared safe.

By choosing the Coulomb gauge, we explicitly wrote down expressions for the dissociation rate of quarkonium, the recombination rate of a heavy quark with an arbitrary heavy antiquark distribution, and the diffusion coefficient of quarkonium. We found that the diffusion coefficient of quarkonium is much smaller than that of the heavy quark. This implies that the dominant energy loss mechanism of quarkonium inside QGP is not diffusion, but rather a sequence of processes: first dissociation, then energy loss as unbound heavy quarks and later recombination. 

The calculations presented here can be generalized to study the effect of a turbulent plasma on quarkonium in the early stage of heavy ion collisions, as is done for heavy quarks \cite{Mrowczynski:2017kso}. For a complete description of quarkonium production in heavy ion collisions, the quarkonium transport equation needs to be coupled with transport equations of heavy quarks. The Boltzmann equations of heavy quarks have been constructed and used in phenomenology \cite{Svetitsky:1987gq,Gossiaux:2008jv,Ke:2018tsh}. By coupling these transport equations, the recombination of quarkonium will be calculated from the real-time dynamical heavy quark distributions rather than phenomenological models. The coupled Boltzmann transport equations have been used to study Upsilon production at both RHIC and LHC and can describe the experimental data \cite{Yao:2018zrg}. In future work, we will solve the coupled Boltzmann equations and study charmonium production in heavy ion collisions.

\begin{acknowledgments}
X.Y. thanks Thomas Mehen and Derek Teaney for helpful discussions and the nuclear theory group at Brookhaven National Laboratory, where part of this work was completed, for its hospitality. The work is supported by U.S. Department of Energy under Research Grant No. DE-FG02-05ER41367. X.Y. also acknowledges support from Brookhaven National Laboratory.
\end{acknowledgments}

\appendix

\bibliographystyle{apsrev4-1}

\end{document}